\documentclass[useAMS,usenatbib]{mn2e}
\usepackage{rotating}
\usepackage{amssymb} 
\usepackage{amsmath} 
\usepackage{graphicx}
\usepackage[nice]{nicefrac}
\usepackage{upgreek}
\usepackage[T1]{fontenc}
\usepackage{ae,aecompl}

\author[Noutsos et al.]
{A. Noutsos$^1$, M. Kramer$^1$, P.~Carr$^3$ and S.~Johnston$^2$\\
$^1$ Max-Planck-Institut f\"ur Radioastronomie, Auf dem H\"ugel 69, 53121 Bonn, Germany.\\
$^2$ Australia Telescope National Facility, CSIRO, P.O. Box 76, 
Epping, NSW 1710, Australia.\\
$^3$ University of Manchester, Jodrell Bank Observatory, Macclesfield, Cheshire, SK11 9DL, UK} 
\date{\today} 
\title[Pulsar Spin--Velocity Alignment]
{Pulsar Spin--Velocity Alignment: Further Results and Discussion}

\def\jnl\style=\rm
\def\ref\jnl#1{{\jnl@style#1}}

\def\aj{\ref@jnl{AJ}}                   
\def\actaa{\ref@jnl{Acta Astron.}}      
\def\araa{\ref@jnl{ARA\&A}}             
\def\apj{\ref@jnl{ApJ}}                 
\def\apjl{\ref@jnl{ApJ}}                
\def\apjs{\ref@jnl{ApJS}}               
\def\ao{\ref@jnl{Appl.~Opt.}}           
\def\apss{\ref@jnl{Ap\&SS}}             
\def\aap{\ref@jnl{A\&A}}                
\def\aapr{\ref@jnl{A\&A~Rev.}}          
\def\aaps{\ref@jnl{A\&AS}}              
\def\azh{\ref@jnl{AZh}}                 
\def\baas{\ref@jnl{BAAS}}               
\def\bac{\ref@jnl{Bull. astr. Inst. Czechosl.}}
\def\caa{\ref@jnl{Chinese Astron. Astrophys.}}
\def\cjaa{\ref@jnl{Chinese J. Astron. Astrophys.}}
\def\icarus{\ref@jnl{Icarus}}           
\def\jcap{\ref@jnl{J. Cosmology Astropart. Phys.}}
\def\jrasc{\ref@jnl{JRASC}}             
\def\memras{\ref@jnl{MmRAS}}            
\def\mnras{\ref@jnl{MNRAS}}             
\def\na{\ref@jnl{New A}}                
\def\nar{\ref@jnl{New A Rev.}}          
\def\pra{\ref@jnl{Phys.~Rev.~A}}        
\def\prb{\ref@jnl{Phys.~Rev.~B}}        
\def\prc{\ref@jnl{Phys.~Rev.~C}}        
\def\prd{\ref@jnl{Phys.~Rev.~D}}        
\def\pre{\ref@jnl{Phys.~Rev.~E}}        
\def\prl{\ref@jnl{Phys.~Rev.~Lett.}}    
\def\pasa{\ref@jnl{PASA}}               
\def\pasp{\ref@jnl{PASP}}               
\def\pasj{\ref@jnl{PASJ}}               
\def\rmxaa{\ref@jnl{Rev. Mexicana Astron. Astrofis.}}%
\def\qjras{\ref@jnl{QJRAS}}             
\def\skytel{\ref@jnl{S\&T}}             
\def\solphys{\ref@jnl{Sol.~Phys.}}      
\def\sovast{\ref@jnl{Soviet~Ast.}}      
\def\ssr{\ref@jnl{Space~Sci.~Rev.}}     
\def\zap{\ref@jnl{ZAp}}                 
\def\nat{\ref@jnl{Nature}}              
\def\iaucirc{\ref@jnl{IAU~Circ.}}       
\def\aplett{\ref@jnl{Astrophys.~Lett.}} 
\def\apspr{\ref@jnl{Astrophys.~Space~Phys.~Res.}}
\def\bain{\ref@jnl{Bull.~Astron.~Inst.~Netherlands}} 
\def\fcp{\ref@jnl{Fund.~Cosmic~Phys.}}  
\def\gca{\ref@jnl{Geochim.~Cosmochim.~Acta}}   
\def\grl{\ref@jnl{Geophys.~Res.~Lett.}} 
\def\jcp{\ref@jnl{J.~Chem.~Phys.}}      
\def\jgr{\ref@jnl{J.~Geophys.~Res.}}    
\def\jqsrt{\ref@jnl{J.~Quant.~Spec.~Radiat.~Transf.}}
\def\memsai{\ref@jnl{Mem.~Soc.~Astron.~Italiana}}
\def\nphysa{\ref@jnl{Nucl.~Phys.~A}}   
\def\physrep{\ref@jnl{Phys.~Rep.}}   
\def\physscr{\ref@jnl{Phys.~Scr}}   
\def\planss{\ref@jnl{Planet.~Space~Sci.}}   
\def\procspie{\ref@jnl{Proc.~SPIE}}   

\makeatletter
\DeclareRobustCommand{\Cpp}
{\valign{\vfil\hbox{##}\vfil\cr
   \textsf{C\kern-.1em}\cr
   $\hbox{\fontsize{\sf@size}{0}\textbf{+\kern-0.05em+}}$\cr}%
}
\makeatother

\begin{document}

\bibliographystyle{mn2e}

\maketitle

\begin{abstract}

The reported alignment between the projected spin-axes and proper motion directions of pulsars is revisited in
the light of new data from Jodrell Bank and Effelsberg. The present investigation uses 54 pulsars, the largest
to date sample of pulsars with proper-motion and absolute polarisation, to study this effect. Our study has
found strong evidence for pulsar spin--velocity alignment, excluding that those two vectors are completely
uncorrelated, with $>99\%$ confidence. Although we cannot exclude the possibility of orthogonal spin--velocity
configurations, comparison of the data with simulations shows that the scenario of aligned vectors is more
likely than that of the orthogonal case. Moreover, we have determined the spread of velocities that a
spin-aligned and spin-orthogonal distribution of kicks must have to produce the observed distribution of
spin--velocity angle offsets. If the observed distribution of spin--velocity offset angles is the result of
spin-aligned kicks, then we find that the distribution of kick-velocity directions must be broad with
$\sigma_v\sim30^\circ$; if the orthogonal-kick scenario is assumed, then the velocity distribution is much
narrower with $\sigma_v\lesssim10^\circ$. Finally, in contrast to previous studies, we have performed
robustness tests on our data, in order to determine whether our conclusions are the result of a statistical
and/or systematic bias. The conclusion of a correlation between the spin and velocity vectors is independent
of a bias introduced by subsets in the total sample.
Moreover, we estimate that the observed alignment is robust to within 10\% systematic uncertainties on the
determination of the spin-axis direction from polarisation data.

\vspace{0.3cm}


\end{abstract}

\section{Introduction}
Pulsars are high-velocity compact stars with transverse speeds ranging from a few tens of km s$^{-1}$ to over
1,000 km s$^{-1}$.
These speeds significantly exceed those of their progenitor stars, which are typically of the order of 10 km
s$^{-1}$ (Lyne \& Lorimer 1994\nocite{ll94}). A number of scenarios have been put forward to explain the high
velocities of pulsars. Classically, an asymmetry in the supernova (SN) explosion --- of the order of 1\% ---
is assumed to provide a kick that imparts the high velocities observed today (Shklovskii
1969\nocite{shk69}). Alternatively, the ``rocket'' model of Tademaru \& Harrison (1975)\nocite{th75}, in which
the pulsar's magnetic-dipole configuration is offset from the progentitor star's centre, can provide birth
velocities of $\sim 100$ km s$^{-1}$. The much higher actual velocities observed and the requirement of this
model for birth spin periods of $P_0\sim 1$ ms have made it less preferred. However, an interesting prediction
of the rocket model is that the pulsars at birth are accelerated in the direction of their spin axes, thus
implying a correlation between the angular and linear momenta.

More recently, Spruit \& Phinney (1998)\nocite{sp98} and Cowsik (1998)\nocite{cow98} proposed a mechanism
whereby a number of impulses imparted onto the pulsar during core collapse provide not only the expected high
birth velocities but also the short birth spin periods ($P_0\sim 10$ ms). In this model, if only a single
impulse of duration $\tau$ is delivered during the collapse, then for $\tau\ll P_0$ this scenario leads to a
spin axis that is preferentially orthogonal to the velocity vector ($\boldsymbol{S}\perp\boldsymbol{v}$);
however, if the pulsar receives a number of such kicks, this correlation is destroyed.
Interestingly, if the the duration of these impulses is of the order of the pulsar's spin period or higher,
then the spin axis is expected to be aligned with the velocity vector
$\boldsymbol{S}\parallel\boldsymbol{v}$); a similar conclusion was reached with toy-model simulations by Wang,
Lai \& Han (2007)\nocite{wlh07}.

Finally, since half of the isolated neutron stars are expected to have come from binary systems, another
mechanism that has been proposed to explain the high pulsar velocities is the break-up of such binary
system during the second SN (Gott, Gunn \& Ostriker 1970\nocite{ggo70}; Bailes 1989\nocite{bai89}). However,
break-up alone cannot account for observed velocities in excess of 1,000 km s$^{-1}$, so a significant
fraction of the observed velocity must also originate from the core collapse phase. At the moment, it seems
that neither the simple conservation of angular momentum nor the binary break-up are enough individually to
impart the observed magnitude of pulsar velocities. Nevertheless, there are alternative physical processes
that have been recently suggested as the trigger mechanism for the presence of asymmetries in the SN ejecta
(Lai et al.~2001\nocite{lcc01}; Janka et al.~2005\nocite{jsk+05}). Those mechanisms can produce fast-moving
pulsars without requiring short birth periods nor very strong magnetic fields.

In addition to the origin of the magnitude of the pulsar velocities, another long-standing issue is the
relative directions of the pulsar velocities to the pulsar spin. As mentioned above, certain SN kick
mechanisms predict a correlation between the spin and velocity vectors of pulsars (e.g.~ Tademaru \& Harisson
1975\nocite{th75}; Spruit \& Phinney 1998\nocite{sp98}). In the case of pulsars whose kinematics and
orientation are the result of the disruption of short-lived, close-binary systems, it has been suggested that
the dominant kick component is due to the impulse received during the break-up of the system: it is therefore
predominantly orthogonal to the spin axis (Colpi \& Wasserman 2003\nocite{cw03b}). A weaker kick component is
associated with a number of internal kicks during the core-collapse process, which tend to align the kick
velocity with the spin axis (Spruit \& Phinney 1998\nocite{sp98}). The above authors present the case of the
double neutron-star system, PSR B1913+16, as an example of the superposition of those two components: Kramer
(1998)\nocite{kra98} used the detection of geodetic precession in that system to measure the misalignment
angle between the pulsar's spin and its orbital momentum (spin tilt); he considered that as direct evidence
for an asymmetric kick during the second SN. Later, Wex et al.~(2000)\nocite{wkk00} concluded that, based on the
orbital dynamics at the time of the second SN explosion, which they constrained using the system's properties
observed at present, the second SN kick must have been directed almost orthogonally to the progenitor's spin.
Similar conclusions were later drawn by Willems et al.~(2004)\nocite{wkh04}, using and updated value of
the spin tilt of PSR B1913+16, which confirmed that it is unlikely that the second SN kick was closely aligned
with the progenitor's spin.

The observational consequence of the above work is that fast pulsars ($\sim 500$ km s$^{-1}$) originating from
close binaries that were disrupted by an asymmetric SN kick --- but whose pre-SN orbital-momentum and spin
vectors were all aligned --- should have space velocities nearly orthogonal to their spin axes. On the other
hand, pulsars with relatively low velocities ($\sim 100$ km s$^{-1}$) --- i.e.~those born in isolated SN
events that imparted a number of weaker kicks on the proto-neutron star --- could predominantly show
spin--velocity alignment.

This distinction between a population of slower pulsars and that of faster ones fits the model of Arzoumanian
et al.~(2002)\nocite{acc02}, who used a bimodal distribution with Gaussian components centred at 90 and 500 km
s$^{-1}$ to describe the distribution of pulsar birth velocities, based on the available sample of 435
isolated Galactic pulsars. However, the recent pulsar-population synthesis by Kuranov, Popov \& Postnov
(2009\nocite{kpp09}; hereafter KPP), who simulated the spins and velocities of pulsars born in both binary
systems and from isolated stars, suggests that the observed distribution of velocities is dominated by the SN
kick mechanism, with the effect of binary progenitors being negligible. This is indeed consistent with the observed
single-mode distribution of 3D velocities, as was derived by Hobbs et al.~(2005)\nocite{hllk05}, from 233
pulsar proper motions. One of the consequences of the kick model used in KPP was that the origin of tightly
aligned spin--velocity configurations (with alignment angles of
$\boldsymbol{S}_{\pm}\angle\boldsymbol{v}<10^\circ$, where $\boldsymbol{S}_\pm$ is the spin orientation) was
mainly found for pulsars born in isolated SNe, having kick velocities of $>50$ km s$^{-1}$. In contrast,
binary progenitors resulted in misaligned configurations, covering the entire range of alignment angles.

Ng \& Romani (2007)\nocite{nr07} performed detailed simulations of the kick dynamics during the core-collapse
event --- independently of any pre-existing systemic velocity (proper/orbital motion) --- and concluded that
fast-moving pulsars are more likely to show spin--velocity correlation than slower-moving ones. The parameters
of their model were determined by fitting a sample of pulsars that included pulsars with (a) available proper
motions but no projected spin-axis orientation, (b) available proper motions and projected spin-axis
orientations (e.g.~from radio-polarimetry measurements), and finally (c) available proper motions and
knowledge of the 3D orientation of their spin axis (from the measured inclination of the associated
pulsar-wind tori). Although a pre-kick spin was required in those simulations in order to produce alignment,
a short initial spin period ($P<20$ ms) was not a strict prerequisite: the simulation produced also aligned
spin--velocity vectors from long-period proto-neutron stars ($P>100$ ms).
Another interesting statistic that was borne out from that study was the standard deviation of the distribution of the 
kick--spin alignment angles, $\sigma_v$, that led to the
observed distribution of kick--spin offsets: the best-determined values were in the range $\sigma_v\approx
10^\circ-20^\circ$. Similar results were derived by the aforementioned population synthesis of KPP; although
it should be noted that comparison with pulsar data by Rankin (2007)\nocite{ran07} led to a narrower
distribution, i.e.~$\sigma_v\approx 5^\circ$--$10^\circ$.

This paper is organised as follows. In Section~\ref{sec:observations}, we present the relevant, observational
background information that needs to be considered in our study. Furthermore, in the same section, we describe
the radio-polarisation observations and the selected data that were used in this paper. In
Section~\ref{sec:dataan}, we perform a statistical analysis on the selected sample and make some concluding
remarks about the observed distribution of spin--velocity angles. In Section 4, we attempt to fit simple
models of the birth-kick distribution to the observed distribution of angles; moreover, we provide estimates
of the width of the angle distributions, for the cases of aligned and orthogonal kicks. At the end of this
section, we perform a systematics check for the robustness of our conclusions against various
degrees of deviation from the published polarisation-angle values; in addition, we try to quantify the
systematic influence of the inclusion of the Johnston et al.~(2005)\nocite{jhv+05} data --- those being the
most convincing evidence for spin--velocity alignment --- in our data set. Finally, in Section 5 we summarise
our results.

\begin{figure*} 
\vspace*{10pt}
\includegraphics[width=1\textwidth]{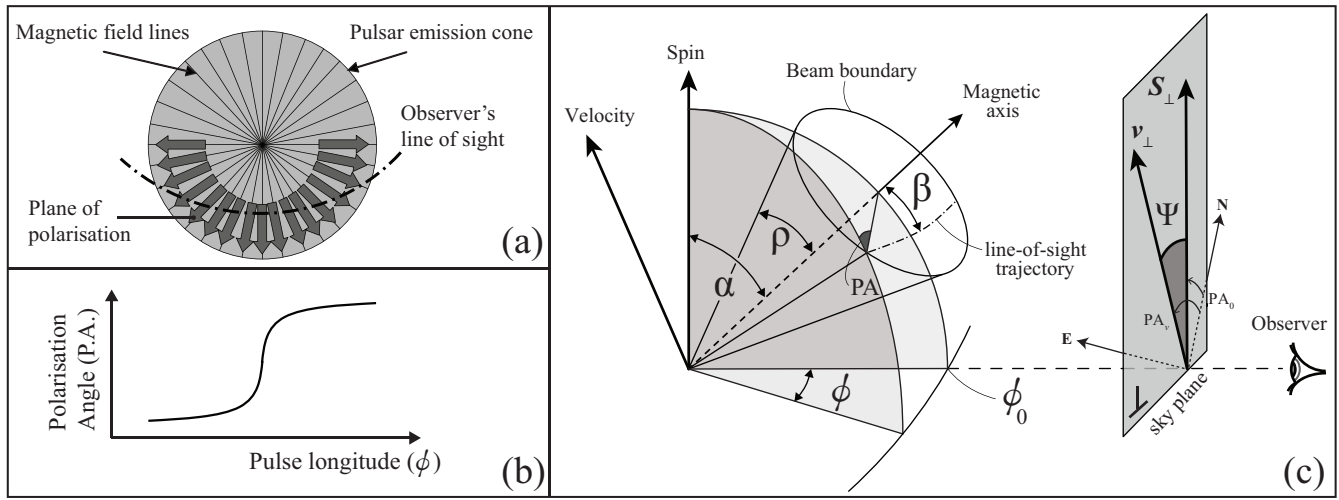}
\caption{\label{fig:geometry1}
 (a) Cross-section of the pulsar beam, showing the sky-projected orientation (radial, solid lines) and
direction (arrows) of the magnetic field lines, and the trajectory of the observer's line of sight
(dash--dotted line), as the beam sweeps past the field of view. The plane of polarisation of the normal mode
of emission at each point of the line of sight is the plane that is perpendicular to the page and contains the
corresponding trace of the magnetic field line. The orthogonal mode of emission occurs perpendicularly to that
plane. (b) The characteristic S-shape of the position angle (PA) rotation across the pulse. According to the
rotating vector model (RVM), as the pulsar beam sweeps the observer, our line-of-sight successively intersects
the different magnetic field lines, corresponding to the different planes of linear polarisation, at different
angles, hence producing the PA-profile shape shown. (c) 3D geometry of the relative configuration between the
spin, magnetic and velocity axes. The beam geometry can be described by two angles: $\alpha$, the angle
between the spin and magnetic axes, and $\beta$, the angle between the magnetic axes and the line-of-sight at
the phase of minimum approach ($\phi_0$). The observer can only determine the direction of the sky-projected
pulsar spin and velocity vectors, $\boldsymbol{S}_\perp$ and $\boldsymbol{v}_\perp$, which are
respectively defined by the position angles ${\rm PA}_0$ and ${\rm PA}_v$, measured north-through-east.
At $\phi_0$, the plane containing the observer and the spin axis also contains the magnetic axis, thus allowing
us to determine the spin-axis orientation. Finally, the opening angle of the emission beam is defined as
$\rho$.}
\end{figure*}

\section{Polarisation Observations}
\label{sec:observations}
\subsection{Geometry}
Ideally, the determination of a pulsar's magnetic-pole orientation, having a north-through-east position angle
(${\rm PA}_0$) on the sky, comes from a rotating-vector-model (RVM) fit to the characteristic `S' shape of the
radio-polarisation position angle (PA) profile. Fig.~\ref{fig:geometry1}a,b shows schematically the geometric
principle behind the rotating vector model, which was developed by Radhakrishnan \& Cooke
(1969)\nocite{rc69a}. Such a model predicts that the point of inflexion in the PA profiles would correspond to
${\rm PA}_0$ and it would therefore be straightforward to determine ${\rm PA}_0$ for any pulsar. However,
there are few cases (e.g.~the Vela pulsar, PSR J2043+2740) where the RVM swing can be clearly detected and
then reliably fitted. Everett \& Weisberg (2001)\nocite{ew01} attempted fits to 70 pulsars' polarisation
profiles and could only reliably fit 10 cases. The main issue is that due to the narrow, conal radio emission,
we can only obtain significant polarisation from pulsars over only a small fraction of their periods, whilst
the rest of the magnetic-field-line rotation remains untraceable. 

As is shown in Fig.~\ref{fig:geometry1}c, the RVM model describes the change of PA as a function of phase, $\phi$, using
these parameters: the angle between the spin and magnetic axes, $\alpha$; the angle between the observer's LOS
and the magnetic axis, $\beta$; and the PA and $\phi$ at the minimum approach of the magnetic pole to the
observer's LOS, PA$_0$ and $\phi_0$, respectively: i.e.~$\tan({\rm PA}-{\rm PA}_0)=\sin\alpha
\sin(\phi-\phi_0)/[\sin(\alpha+\beta)\cos\alpha-\cos(\alpha+\beta)\sin\alpha\cos(\phi-\phi_0)]$.
For those pulsars where the RVM model can satisfactorily describe the observed PA profile, the relative orientation between the pulsar spin, magnetic axis and the observer's line of sight (LOS) can be derived.

Despite the limitations of RVM fitting, if one is only interested in determining ${\rm PA}_0$, there are
alternative arguments that can help: e.g.~one can use other indicators, like the pulse profile's centre of
symmetry or the change of handedness of the circular-polarisation profile (see Section~\ref{ssubsec:rvmrel}
below). The latter feature, when it occurs near the profile's peak, has been associated with emission from
near the magnetic pole and viewing angles almost along the local magnetic field (Rankin 1986\nocite{ran86}).
In the present work, when RVM fitting alone was deemed unreliable with our data, we also considered arguments
based on pulsar symmetry, which took into account the distribution and size of the profile's peaks, as was
done in Johnston et al.~(2005, 2007). Such geometrical investigation benefited from
available multi-frequency profiles for the pulsars in our sample, in order to account for a possible range in
emission heights (Kramer et al.~1997\nocite{kxj+97}). Overall, in most cases the determined ${\rm PA}_0$ was a
product of all the above considerations.

It is also worth acknowledging that any of the above approaches of determining ${\rm PA}_0$ may potentially
lead to erroneous results due to aberration/retardation (A/R) effects, which, although small, can potentially
cause the inflexion point of the RVM swing to appear delayed with respect to the pulse profile's mid-point
(Blaskiewicz et al.~1991\nocite{bcw91}; Dyks et al.~2004\nocite{drh04}). In addition, A/R effects can shift
the positions of conal components to earlier observed phases, thus misleading arguments based on profile
symmetry. Clearly, observations of rapidly-rotating pulsars will be more prone to the above effects. Recently,
Krzeszowski et al.~(2009)\nocite{kmg+09} --- and earlier Mitra \& Li (2004)\nocite{ml04} --- investigated the
presence of A/R effects in pulsar profiles from multi-frequency observations. Amongst their conclusions was
that the peak emission does not coincide with the fiducial phase containing the magnetic and spin axes and that,
overall, the measured phase lags due to A/R were of the order of 1 deg.

Upon measuring ${\rm PA}_0$, which is always measured relative to the orthogonal frame defined by our
polarisation feeds, we have effectively determined an orthogonal reference frame on the sky plane that defines
the spin-axis orientation. This becomes clear if we assume that the plane of polarisation of the measured
signal (defined by PA) coincides with the plane defined by the magnetic-field lines and the magnetic axis, as
is shown in Fig.~\ref{fig:geometry1}c. At the phase of minimum approach of the observer's LOS to the magnetic
pole (i.e.~at ${\rm PA}_0$), that plane should also contain the spin axis. So, in that way, we can say that we
have determined the spin-axis orientation, i.e.~${\rm PA}_{\rm spin}\equiv{\rm PA}_0$. Note that the
calculation of PA from the Stokes $Q$ and $U$ can only provide a `headless' polarisation vector,
$P=L\exp[\tan^{-1}(U/Q)/(2i)]$, introducing therefore a 180$^\circ$ ambiguity. In addition, as was
shown by Backer, Rankin \& Campbell (1975)\nocite{brc75} and Manchester, Taylor \& Huguenin (1975)\nocite{mth75},
pulsar emission can occur in two orthogonal modes: i.e.~the polarisation plane can either coincide with the
field line--magnetic axis plane or be perpendicular to it. Hence, there is an additional 90$^\circ$ ambiguity
in the projected direction of the spin axis. And so, given the aforementioned ambiguities, the angle
between the spin axis and the velocity vector, $\Psi$, as is measured on the sky plane (refer to
Fig.~\ref{fig:geometry1}c), can only be unambiguously given in the interval $0^\circ$--$45^\circ$.

Observationally, there are a number of obstacles that prevent us from recovering the intrinsic distribution of
spin--velocity angles. A major limitation is the unknown velocities of the progenitor systems (whether
isolated or binary), which, if high, can smear even a perfect correlation to an apparent misalignment. Very
importantly, any study of the subject has to rely on the projected velocities and spin-axes directions;
without knowledge of the pulsar's orientation and motion along the LOS, any evidence for
alignment may be coincidental. There are indeed cases where the inclination of pulsar-wind tori can reveal the
3D orientation of the pulsar's spin (e.g.~in the case of the Crab and Vela Pulsar Wind Nebulae --- PWNe; Ng \& Romani
2004\nocite{nr04}), but even in those cases the unknown progenitor velocity can dominate over the measurement
errors of the proper motion, depending on the distance to the SN site, thus leading to significant
uncertainties on the alignment angles (for a discussion, see \S3.2 of Kaplan et al.~2008\nocite{kcga08}).

\subsubsection{The reliability of RVM fits}
\label{ssubsec:rvmrel}
There are several examples of pulsars whose profile characteristics allow one to check the consistency of the
derived value of PA$_0$ from an RVM fit with that based on e.g.~pulse symmetry, evolution with frequency and
features in circular polarisation. A few clear examples can be mentioned from the work of Johnston et
al.~(2005, 2007); the reader can refer to Fig.~\ref{fig:fig1}, for the polarisation
profiles of the pulsars mentioned here. Success cases of $\phi_0$ determination through RVM fitting include
that of PSR J1740+1311, which has a five-component profile with the RVM inflexion point coinciding with the
profile's centre; similarly, PSR J2048$-$1616 has a well-defined RVM swing across its triple-component
profile, at all examined frequencies, with the inflexion point coinciding with mid-point between the outrider
components.

However, there are many cases for which the RVM model does not provide a clear determination of $\phi_0$,
often due to orthogonal-mode emission and other features that deviate from an expected RVM swing; in those
cases, e.g.~PSR J1735$-$0724, the profile characteristics may provide alternative hints at the location of the minimum
approach to the magnetic pole. The aforementioned pulsar exhibits a clear peak at the location of the central
component, where $\phi_0$ is chosen, but RVM fitting results in large uncertainties due to the presence of a
hump in the PA profile (most likely caused by a transition, a `jump', between normal- and orthogonal-mode
emission); in several cases, the inclusion of orthogonal jumps at preselected locations in the RVM model helps
towards the determination. Also, it was mentioned earlier that the change of handedness of the circular
polarisation is telling of emission from near the magnetic pole. A vivid example is that of PSR J1900$-$2600,
for which the RVM inflexion, the profile's mid-point {\em and} the change of handedness of the circular
polarisation all coincide; thus, for this pulsar, the confidence in the determined value of PA$_0$ is
amongst the highest presented here.

Interestingly, there are cases for which the RVM model, even with the inclusion of orthogonal
jumps, yields a different solution to that suggested by profile symmetry, etc. A good example is that of PSR
J0738$-$4042, for which a very good RVM solution for $\phi_0$ trails the location of the profile's mid-point
by $\approx 4^\circ$; also, a lag of $2^\circ$ between the profile's mid-point and the well-determined RVM
inflexion is observed for PSR J0452$-$1759. Lags of that magnitude may be associated with A/R effects, although there are cases where this association is tentative (Gupta \& Gangadhara 2003\nocite{gg03}). 

Finally, it is also worth noting those pulsars whose PA profiles are completely featureless (i.e.~flat)
across the pulse; for such cases, like that of PSR J1915+1009, which has a single component, RVM fitting does
not help, and profile features like the location of the peak intensity are normally employed. However, it
should be cautioned that such determination assumes that the observable emission corresponds to a core
component, i.e.~emanating from near the magnetic pole. If this is not the case, the determined $\phi_0$ and
the corresponding PA$_0$ can be completely wrong.

\subsection{Previous Work}
\label{subsec:prevwork}
Amongst the recent noteworthy observational investigations of pulsar spin--velocity is that of Johnston et
al.~(2005; hereafter J05\nocite{jhv+05}) who used 25 carefully calibrated pulsars, 10 of which showed strong
evidence for a correlation between spin and velocity. The uniform distribution of $\Psi$ between $0^\circ$ and
$45^\circ$ (i.e.~allowing for orthogonal emission) was rejected under a Kolmogorov--Smirnov (KS) test at the
$98\%$ confidence level (CL). It should be mentioned that the expected form of the $\Psi$ distribution for
${\rm PA}_0 \perp {\rm PA}_v$ (see Section \ref{sec:toymodel}) was also rejected at the 99\% CL. In addition,
for those pulsars in that sample with spin-down ages, $\tau_{\rm c}=P/(2\dot{P})<3$ Myr, where $\dot{P}$ is
the period time-derivative, the authors found that $\Psi<10^\circ$, although it should be stated that the
sample was only composed of 4 pulsars.

The follow up work of Johnston et al.~(2007)\nocite{jkk+07} used a different sample of 22 pulsars, only 7 of
which were found to be plausibly aligned. The conclusion was that the evidence for alignment found in the
first study is not supported by the separate, follow-up sample. However, it was stressed that the pulsars in
the second study were both older and more distant than those of the first and, therefore, could be more prone
to the effect of the gravitational potential on the velocity direction.

In an attempt to assess the evidence for spin-velocity alignment in view of better statistics, Rankin
(2007)\nocite{ran07} re-examined the sample of Johnston et al.~(2005)\nocite{jhv+05} and revised the ${\rm
PA}_0$ values for 23 pulsars of the original sample by studying the individual profiles' polarisation and, in
particular, their frequency evolution. Additionally, Rankin included polarisation measurements from
Karastergiou \& Johnston (2006)\nocite{kj06}, Xilouris et al.~(1991)\nocite{xrss91} and Morris et al.~(1979,
1981)\nocite{mgs+79}\nocite{mgs+81}. Finally, a small number of additional measurements were added to the
above sample from new L-band observations with Arecibo, amounting to a total of 46 values of ${\rm PA}_0$ and
the corresponding ${\rm PA}_v$ values from Hobbs et al.~(2005)\nocite{hllk05}. Interestingly, in this work,
the offset angle ${\rm PA}_0-{\rm PA}_v$ was not constrained between 0$^\circ$ and 90$^\circ$ or 0$^\circ$ and 45$^\circ$, thus not
folding the polarisation-measurement or orthogonal-emission ambiguities into the final result. Nevertheless,
the distribution of ${\rm PA}_0-{\rm PA}_v$ (between 0 and 180$^\circ$) of the total pulsar sample showed
an evident clustering of values around 0$^\circ$ and 90$^\circ$. Hence, it was argued, based on the majority
of cases being nearly aligned or nearly orthogonal, that there is convincing evidence for a
correlation between the spin and velocity directions of pulsars.

Crucially, despite the preferential concentration of values around 0$^\circ$ and 90$^\circ$, in the above
studies it was recognised that the results are inconclusive as to what SN-kick mechanism is in operation or
whether, indeed, the distinction between orthogonal or aligned configurations can be trusted due to
orthogonal-mode emission.

Finally, even though the data revision by Rankin (2007) did not contradict, but rather strengthened, the
conclusions of Johnston et al.~(2005), it highlighted the subjectivity of determining ${\rm PA}_0$
through studies of pulsar PA profiles. It is therefore fair to say that, ultimately, without the deeper
understanding of pulsar magnetospheric physics, the conclusions from any such analysis will unavoidably be
subject to differing interpretations. This is clearly the weak point of all polarisation-based analyses to
date. Recognising that our conclusions may potentially be governed by a bias in the selection of ${\rm PA}_0$,
in Section~\ref{sec:sys} we explore the dependence of the distribution of $\Psi$ on various degrees of deviation
from the published values of ${\rm PA}_0$.

\begin{table*}
\caption{\label{tab:tab1} 
Table of velocity and spin orientation for the 54 pulsars used in our study: Column 3 shows the position angle
of the projected velocity vector, ${\rm PA}_v$, derived from pulsar timing and VLBI (Very Long Baseline Interferometry) observations; Column 5
lists the position angle of the spin-axis projection onto the sky plane, ${\rm PA}_0$, as was derived from
radio polarisation measurements and fits to the PWN X-ray images; finally, Column 7 shows the minimum offset
angle between projected velocity and spin, $\Psi=({\rm PA}_0-{\rm PA}_v) \mod 45^\circ$, which accounts for
polarisation and orthogonal-mode ambiguities. The statistical errors on the ${\rm PA}_v$, ${\rm PA}_0$ and
$\Psi$ are shown in parentheses and refer to the last significant digit. More information on the observations
and the respective techniques used can be found in the publications referenced in Columns 4 and 6.} \centering
\begin{normalsize}
\begin{tabular}{@{}lrrrrrr}
\\
\hline 
PSR   &  $\log\tau_{\rm c}$  &    ${\rm PA}_v$ \ \  &  Reference  &  ${\rm PA}_{0}$ \ \        &   Reference        &  $\Psi$             \\ 
      &  [yr]                &    [$^\circ$]  \ \         &             &  [$^\circ$]  \ \     &                    &  [$^\circ$]   \\ 
\hline

J0139+5814              &          5.6            &       30(13)            \  \            &          [1]   \hspace{0.7cm}         &     $-$71(5)            \  \            &         [18]   \hspace{0.7cm}        &       11(14)           \\
J0152$-$1637            &          7.0            &       173(3)            \  \            &          [5]   \hspace{0.7cm}         &        92(9)            \  \            &         [19]   \hspace{0.7cm}         &         8(9)           \\
J0304+1932              &          7.2            &     $-$9(11)            \  \            &          [2]   \hspace{0.7cm}         &       12(10)            \  \            &         [18]   \hspace{0.7cm}         &       21(15)           \\
J0332+5434              &          6.7            &     $-$61(1)            \  \            &          [3]   \hspace{0.7cm}         &      $-$4(5)            \  \            &         [18]   \hspace{0.7cm}         &        33(5)           \\
J0358+5413              &          5.8            &       69(16)            \  \            &          [4]   \hspace{0.7cm}         &     $-$33(5)            \  \            &         [18]   \hspace{0.7cm}         &       12(17)           \\
J0452$-$1759            &          6.2            &       72(23)            \  \            &          [5]   \hspace{0.7cm}         &        47(3)            \  \            &         [20]   \hspace{0.7cm}         &       25(23)           \\
J0454+5543              &          6.4            &     $-$72(3)            \  \            &          [1]   \hspace{0.7cm}         &    $-$66(11)            \  \            &         [18]   \hspace{0.7cm}         &        6(11)           \\
J0525+1115              &          7.9            &      132(16)            \  \            &          [5]   \hspace{0.7cm}         &     $-$65(4)            \  \            &         [21]   \hspace{0.7cm}         &       17(16)           \\
J0534+2200              &          3.1            &      292(10)            \  \            &          [6]   \hspace{0.7cm}         & 124.0(1)$^{\rm b}$                      &         [22]   \hspace{0.7cm}         &       12(10)            
\\             
J0538+2817              &          4.6            &   $-$24.1(2)            \  \            &          [7]   \hspace{0.7cm}         &    $-$49(10)            \  \            &         [18]   \hspace{0.7cm}         &       25(10)           
\\
J0543+2329              &          5.4            &       57(20)            \  \            &          [1]   \hspace{0.7cm}         &    $-$100(5)            \  \            &         [18]   \hspace{0.7cm}         &       23(20)           \\
J0630$-$2834            &          6.4            &       294(3)            \  \            &          [8]   \hspace{0.7cm}         &        26(2)            \  \            &         [21]   \hspace{0.7cm}         &         2(4)           \\
J0659+1414              &          5.0            &      93.1(4)            \  \            &          [9]   \hspace{0.7cm}         &     $-$86(2)            \  \            &         [20]   \hspace{0.7cm}         &         1(2)           \\
J0738$-$4042            &          6.6            &       313(5)            \  \            &          [8]   \hspace{0.7cm}         &     $-$21(2)            \  \            &         [20]   \hspace{0.7cm}         &        26(5)           \\
J0742$-$2822            &          5.2            &       278(5)            \  \            &          [4]   \hspace{0.7cm}         &   $-$81.7(1)            \  \            &         [21]   \hspace{0.7cm}         &         0(5)           \\
J0820$-$1350            &          7.0            &       159(6)            \  \            &          [5]   \hspace{0.7cm}         &        65(2)            \  \            &         [21]   \hspace{0.7cm}         &         4(6)           \\
J0826+2637              &          6.7            &     $-$34(1)            \  \            &          [2]   \hspace{0.7cm}         &       36(10)            \  \            &         [18]   \hspace{0.7cm}         &       20(10)           \\
J0835$-$4510            &          4.1            &     301.0(1)            \  \            &         [10]   \hspace{0.7cm}         &      36.8(1)            \  \            &         [21]   \hspace{0.7cm}         &       5.8(1)           \\
J0837+0610              &          6.5            &         2(5)            \  \            &          [5]   \hspace{0.7cm}         &        18(5)            \  \            &         [20]   \hspace{0.7cm}         &        16(7)           \\
J0837$-$4135            &          6.5            &       187(6)            \  \            &          [8]   \hspace{0.7cm}         &     $-$84(5)            \  \            &         [20]   \hspace{0.7cm}         &         1(8)           \\
J0922+0638              &          5.7            &        12.0(1)            \  \            &         [11]   \hspace{0.7cm}         &     $-$56(5)            \  \            &         [18]   \hspace{0.7cm}         &        22(5)           \\
J0953+0755              &          7.2            &     355.9(2)            \  \            &          [3]   \hspace{0.7cm}         &      14.9(1)            \  \            &         [21]   \hspace{0.7cm}         &      19.0(2)           \\
J1136+1551              &          6.7            &     348.6(1)            \  \            &          [3]   \hspace{0.7cm}         &     $-$78(2)            \  \            &         [21]   \hspace{0.7cm}         &        23(2)           \\
J1239+2453              &          7.4            &     295.0(1)            \  \            &          [3]   \hspace{0.7cm}         &     $-$66(1)            \  \            &         [21]   \hspace{0.7cm}         &         1(1)           \\
J1430$-$6623            &          6.7            &       236(9)            \  \            &         [12]   \hspace{0.7cm}         &   $-$28.5(7)            \  \            &         [21]   \hspace{0.7cm}         &         6(9)           \\
J1453$-$6413            &          6.0            &       217(3)            \  \            &         [12]   \hspace{0.7cm}         &   $-$56.9(4)            \  \            &         [21]   \hspace{0.7cm}         &         4(3)           \\
J1456$-$6843            &          7.6            &     252.7(6)            \  \            &         [13]   \hspace{0.7cm}         &   $-$31.6(6)            \  \            &         [21]   \hspace{0.7cm}         &      14.3(8)           
\\
J1509+5531              &          6.4            &        47(2)            \  \            &          [2]   \hspace{0.7cm}         &      $-$3(7)            \  \            &         [18]   \hspace{0.7cm}         &        40(7)           \\
J1604$-$4909            &          6.7            &       268(6)            \  \            &         [13]   \hspace{0.7cm}         &     $-$17(3)            \  \            &         [20]   \hspace{0.7cm}         &        15(7)           \\
J1645$-$0317            &          6.5            &       353(3)            \  \            &          [8]   \hspace{0.7cm}         &        56(4)            \  \            &         [21]   \hspace{0.7cm}         &        27(5)           \\
J1709$-$1640            &          6.2            &      192(16)            \  \            &          [5]   \hspace{0.7cm}         &        15(2)            \  \            &         [21]   \hspace{0.7cm}         &        3(16)           \\
J1709$-$4429            &          4.2            &      160(10)$^{\rm a}$                       &         [14]   \hspace{0.7cm}         & 163.6(7)$^{\rm b}$                      &         [23]   \hspace{0.7cm}         &         
4(10)          \\   
J1735$-$0724            &          6.7            &       355(3)            \  \            &          [8]   \hspace{0.7cm}         &        55(5)            \  \            &         [20]   \hspace{0.7cm}         &        30(6)           \\
J1740+1311              &          6.9            &       228(6)            \  \            &          [8]   \hspace{0.7cm}         &     $-$46(4)            \  \            &         [21]   \hspace{0.7cm}         &         4(7)           \\
J1801$-$2451            &          4.2            &       270$^{\rm c}$    \                &         [15]   \hspace{0.7cm}         &     $-$55(5)            \  \            &         [20]   \hspace{0.7cm}         &        35(5)           
\\
J1820$-$0427            &          6.2            &      338(17)            \  \            &          [5]   \hspace{0.7cm}         &        42(3)            \  \            &         [20]   \hspace{0.7cm}         &       26(17)           \\
J1844+1454              &          6.5            &       36(15)            \  \            &          [5]   \hspace{0.7cm}         &     $-$52(2)            \  \            &         [21]   \hspace{0.7cm}         &        2(15)           \\
J1850+1335              &          6.6            &      237(16)            \  \            &          [5]   \hspace{0.7cm}         &     $-$45(3)            \  \            &         [20]   \hspace{0.7cm}         &       12(16)           \\

\hline 
\end{tabular} 
\end{normalsize}

\end{table*}

\nocite{hla93}
\nocite{las82}
\nocite{bbgt02}
\nocite{fgm+97}
\nocite{hllk05}
\nocite{cm99}
\nocite{nrb+07}
\nocite{bfg+03a}
\nocite{btgg03b}
\nocite{dlrm03}
\nocite{ccl+01}
\nocite{bmk+90a}
\nocite{bmk+90b}
\nocite{wlh06}
\nocite{bgc+06}
\nocite{car07}
\nocite{nr04}
\nocite{rndb05}
\nocite{fgbc99}
\nocite{mgb+02}

\setcounter{table}{0}

\begin{table*}
\caption{\label{tab:tab1} Continued.}
\centering
\begin{normalsize}
\begin{tabular}{@{}lrrrrrr}
\\
\hline 
PSR   &  $\log\tau_{\rm c}$  &    ${\rm PA}_v$  &  Reference  &  ${\rm PA}_{0}$  \ \  &   Reference     & $\Psi$               \\ 
      &  [yr]                &    [$^\circ$]          &             &  [$^\circ$]  \ \     &            &     [$^\circ$]       \\ 
\hline

J1900$-$2600            &          7.7            &     202.8(7)            &         [16]  \hspace{0.7cm}          &     $-$43(2)            \ \            &         [21]  \hspace{0.7cm}          &        24(2)           \\
J1913$-$0440            &          6.5            &      166(11)            &          [5]  \hspace{0.7cm}          &     $-$68(2)            \ \            &         [21]  \hspace{0.7cm}          &       36(11)           \\
J1915+1009              &          5.6            &      174(15)            &          [5]  \hspace{0.7cm}          &        85(3)            \ \            &         [20]  \hspace{0.7cm}          &        1(15)           \\
J1921+2153              &          7.2            &       34(12)            &          [5]  \hspace{0.7cm}          &     $-$35(2)            \ \            &         [21]  \hspace{0.7cm}          &       21(12)           \\
J1932+1059              &          6.5            &      65.2(2)            &         [11]  \hspace{0.7cm}          &   $-$11.3(1)            \ \            &         [21]  \hspace{0.7cm}          &      13.5(2)           \\
J1935+1616              &          6.0            &       176(1)            &          [5]  \hspace{0.7cm}          &      10.1(7)            \ \            &         [21]  \hspace{0.7cm}          &        14(1)           \\
J1937+2544              &          6.7            &       220(9)            &          [5]  \hspace{0.7cm}          &      $-$9(5)            \ \            &         [20]  \hspace{0.7cm}          &       41(10)           \\
J1952+3252              &          5.0            &       252(7)            &         [17]  \hspace{0.7cm}          &  85(5)$^{\rm b}$                       &         [22]  \hspace{0.7cm}          &       13(9)            \\
J1955+5059              &          6.8            &     $-$27(3)            &          [5]  \hspace{0.7cm}          &       19(10)            \ \            &         [18]  \hspace{0.7cm}          &       44(10)           \\
J2013+3845              &          5.6            &       47(11)            &          [5]  \hspace{0.7cm}          &        83(5)            \ \            &         [18]  \hspace{0.7cm}          &       36(12)           \\
J2018+2839              &          7.8            &        23(2)            &          [3]  \hspace{0.7cm}          &       89(20)            \ \            &         [18]  \hspace{0.7cm}          &       24(20)           \\
J2022+2854              &          6.5            &        11(1)            &          [3]  \hspace{0.7cm}          &       12(10)            \ \            &         [18]  \hspace{0.7cm}          &        1(10)           \\
J2022+5154              &          6.4            &   $-$24.5(9)            &          [3]  \hspace{0.7cm}          &        13(5)            \ \            &         [18]  \hspace{0.7cm}          &        38(5)           \\
J2048$-$1616            &          6.5            &        92(2)            &          [4]  \hspace{0.7cm}          &     $-$13(5)            \ \            &         [20]  \hspace{0.7cm}          &        15(5)           \\
J2157+4017              &          6.8            &     $-$81(3)            &          [1]  \hspace{0.7cm}          &       79(10)            \ \            &         [18]  \hspace{0.7cm}          &       21(10)           \\
J2330$-$2005            &          6.7            &        86(2)            &          [8]  \hspace{0.7cm}          &       21(10)            \ \            &         [20]  \hspace{0.7cm}          &       25(10)           \\

\hline 
\end{tabular} 
\end{normalsize}

\flushleft {\sc References} ---  Proper motions: [1]~Harrison et al.~(1993); \, [2]~Lyne et al.~(1982); 
\, [3]~Brisken et al.~(2002); \, [4]~Fomalont et al.~(1997); \, [5]~Hobbs et al.~(2005);  
\, [6]~Caraveo \& Mignani~(1999);  \, [7]~Ng et al.~(2007);\,  [8]~Brisken et al.~(2003a);
\, [9]~Brisken et al.~(2003b);  \, [10]~Dodson et al.~(2003);
\, [11]~Chatterjee et al.~(2001);  \, [12]~Bailes et al.~(1990b);
\, [13]~Bailes et al.~(1990a);  \, [14]~Wang et al.~(2006); \, [15]~Blazek et al.~(2006);
\, [16]~Formalont et al.~(1999); \, [17]~Migliazzo et al.~(2002).
\ Polarisation: [18]~Carr~(2007); \, [19]~Wang et al.~(2006);  \, [20]~Johnston et al.~(2007);
\, [21]~Johnston et al.~(2005);  \, [22]~Ng \& Romani (2004); \, [23]~Romani et al.~(2005).

\vspace{0.3cm}

a.~Assumes an association between the pulsar and the SNR G343.1$-$2.3 (see e.g.~Dodson \& Golap~2002\nocite{dg02}). \\
b.~Spin axis direction derived from fitting the tori of the PWN seen in {\em Chandra} images.\\
c.~Position angle corresponding to the upper limit on the westerly motion of this pulsar, derived from interferometric observations. 

\end{table*}

 \begin{figure*} 
\vspace*{10pt}
\includegraphics[width=0.9\textwidth]{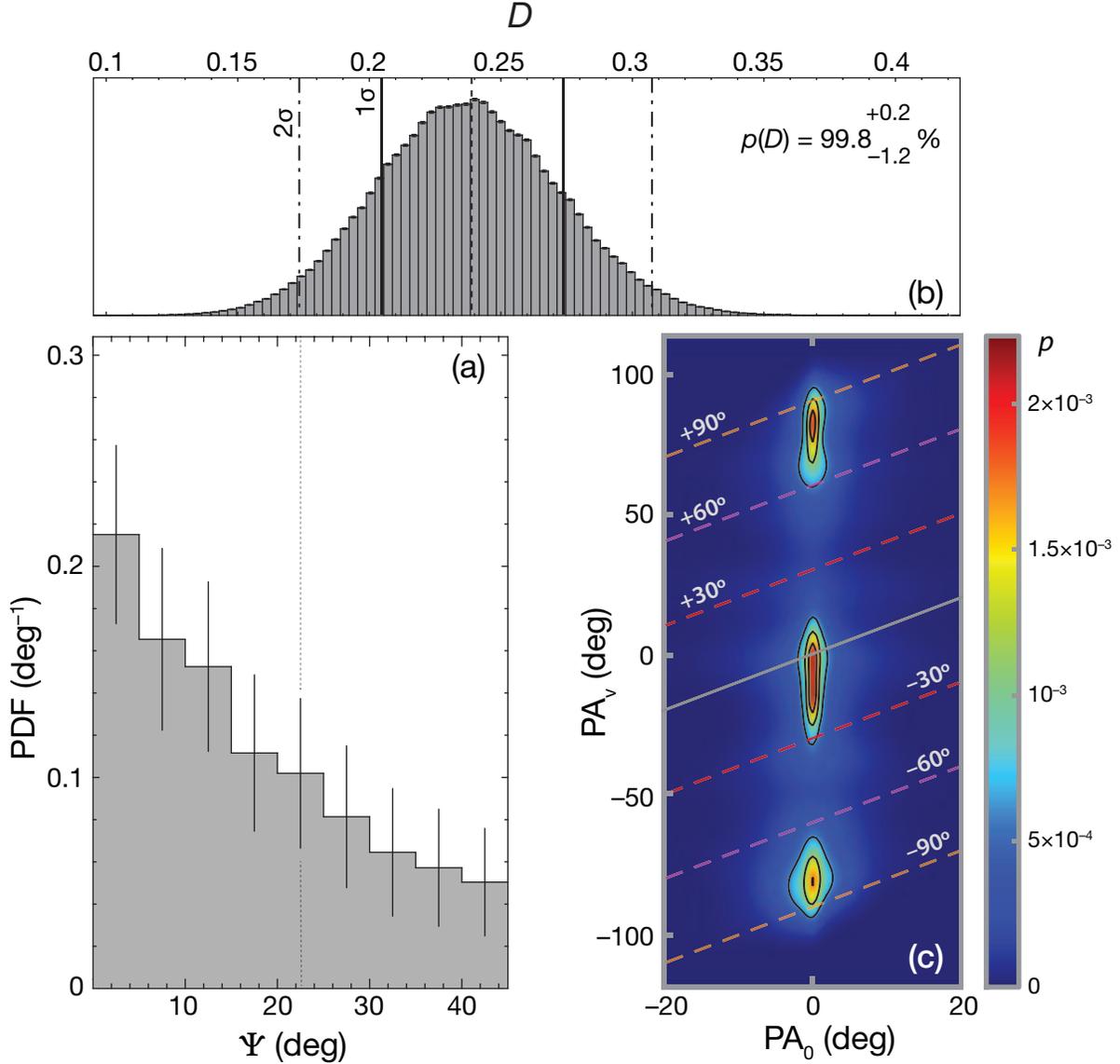}
\caption{\label{fig:psihist} 
(a) Distribution of $\Psi$, for the pulsars of Table~\ref{tab:tab1}, from $10^6$ Monte Carlo simulations of
${\rm PA}_0$ and ${\rm PA}_v$, based on their published error values. The vertical, dotted grey line
marks the mid-range of the $\Psi$ parameter space ($\Psi=22\fdg.5$). (b) Distribution of the KS statistic,
$D$, from the above simulations, from the comparison of the distribution shown in (a) with a uniform
distribution. The probability of rejecting the uniform distribution, $p(D)$, and the 1$\sigma$ confidence
interval is also shown. (c) Two-dimensional PDF of ${\rm PA}_v$ and ${\rm PA}_{0}$, generated from
random Gaussian iterations of the values in our data within their respective errors. To produce this map, we re-defined
${\rm PA}_v$ as ${\rm PA}_v-{\rm PA}_0 \mod 90^\circ$, considering only the polarisation
ambiguity, and forced ${\rm PA}_0$ to 0. The final distribution shown was generated by randomising the the
re-defined ${\rm PA}_v$ and ${\rm PA}_0$ within their errors and wrapping their difference modulo $90^\circ$, when
necessary. The solid grey and dashed colour lines overlaid on the map correspond to the $0^\circ, \pm30^\circ,
\pm 60^\circ$ and $\pm 90^\circ$ spin--velocity offset contours. The colour scale indicates the
probability, $p$, increasing from blue towards red with graduated colours, corresponding to the different
points on the map.}
\end{figure*}

\subsection{Data Sample}
\label{subsec:datasel}
For this paper, we have tried to compile a database of all previously published measurements of ${\rm PA}_0$
that are accompanied by known measurements of ${\rm PA}_v$ from the literature. The aim of this paper is to
provide a conclusive statement on whether the observed, 2-dimensional spin and velocity orientations of
pulsars are correlated, for a sample of pulsars with reliable, absolutely-calibrated polarisation
measurements. Although we considered all published measurements of ${\rm PA}_0$ and ${\rm PA}_v$, we only
included in our analysis those measurements for which published polarisation profiles were available and whose
quality could be examined. The final sample that was used in this work contained 51 pulsars, whose
polarisation profiles have allowed the determination of spin-axis directions, and an additional 3 pulsars,
whose spin-axis directions have been determined through fitting the the tori of the PWNe seen in {\em Chandra}
X-ray images (see Table~\ref{tab:tab1} for references).

The pulsar polarisation profiles used in this work came from observations that were performed
with the Lovell and Effelsberg radio telescopes as well as from older observations with the Parkes telescope.
The details concerning the Parkes observations and data selection can be found in the published articles by
Johnston et al.~(2005, 2007)\nocite{jhv+05}\nocite{jkk+07} and Wang et al.~(2006)\nocite{wlh06}.
The Lovell observations were carried out in three sessions: 2005, 18 September; 2006, 23 March; and 2006, 30
May. The centre frequency of the observations was 1.4 GHz; the 21-cm cryogenic receiver installed at the focus
of the Lovell telescope is sensitive to both senses of circular polarisation and has a system equivalent flux
density (SEFD) of 37 Jy. The data were dedispersed coherently and in real time using the known pulsar dispersion measure (DM) and the COBRA (Coherent On-line Baseband Receiver for Astronomy) coherent-dedispersion back end (\nocite{car07}Carr 2007). Our observations with this backend used different
bandwidths between 10 and 30 MHz, although the COBRA hardware itself is capable of coherently dedispersing up
to 100 MHz. Depending on the pulsar luminosity, the integration time of those observations varied between 3
and 15 minutes. A total of 12 pulsars were observed with the Lovell system.

The Effelsberg observations were performed at two frequencies: 2.7 GHz and 4.85 GHz, utilising 80 and 500 MHz
of bandwidth, respectively. The 2.7-GHz receiver is equipped with circular-polarisation feeds and provides an
SEFD of 11 Jy; the 4.85-GHz receiver also has circular-polarisation feeds and an SEFD of 18 Jy. The
observations took place during 2006, 12--14 March and 2006, 27 May, during which a total of 32 pulsars were
observed, with 21 of these being different from those observed with the Lovell. The data were acquired and
folded with the Effelsberg Pulsar Observation System (EPOS) and the PUB86 pulsar back end
(\nocite{kra95}Kramer 1995). Dedispersion was not necessary, as the dispersive delay for the observed pulsars,
given the available bandwidth at the relevant observing frequencies, was negligible compared the pulse period:
i.e.~the dispersive delay between two frequencies, $f_1$ and $f_2$ (in MHz), for a pulsar with a dispersion
measure DM (in cm$^{-3}$pc), is given by $\Delta t=(4.15\times 10^6 \ {\rm ms})\times {\rm DM}\times
(f_1^{-2}-f_2^{-2})$; for a bandwidth of 30 MHz at 2.7 GHz and at 4.85 GHz, this delay is $\approx 12$ $\mu$s
and $\approx 2$ $\mu$s per DM unit, respectively. The pulsars in our sample have ${\rm DM}\lesssim 100$
cm$^{-3}$pc and $P\gtrsim 0.2$ s, which means that dispersion smearing is confined to, at most, only $\approx
0.5\%$ of the pulse period.

In both the Lovell and Effelsberg polarisation observations, the Stokes data of each observation were
absolutely calibrated by correcting, amongst other effects, for the parallactic-angle rotation during the
observation and the Faraday rotation, using the published rotation measure (RM) of the pulsar
(\nocite{mhth05}Manchester et al.~2005). In general, the error on RM, $\sigma_{\rm RM}$, could significantly
impact on the PA$_0$ error, $\sigma_{\rm PA}$; the latter is given by $\sigma_{\rm PA}=c^2\sigma_{\rm
RM}/f^2$, where $c$ is the speed of light and $f$, the observing frequency. In Johnston et al.~(2005, 2007), the RM
errors were folded into the final, published uncertainty on ${\rm PA}_0$; this was not done for the pulsars of
observed with the Lovell and Effeslberg telescopes. However, for the vast majority of those pulsars
$\sigma_{\rm RM}<1$ rad m$^{-1}$, which at the observing frequencies in question meant that $\sigma_{\rm
PA}\lesssim 1^\circ$. More importantly, for all those pulsars the measurement errors on ${\rm PA}_v$ and
${\rm PA}_0$ dominate over the errors due to RM uncertainties:
the dominant measurement errors are due to (a) the limited signal-to-noise ratio (S/N) in linearly polarised intensity, $L$,
i.e.~$\sigma_{\rm PA}\sim \sigma_I/L$, where $\sigma_I$ is the rms of the total intensity; and (b) due to
uncertainties in proper motion determination. The worst-case scenario in our sample is PSR J0922+0638,
observed with Effelsberg at 2.7 GHz: this pulsar has ${\rm RM}=32\pm 2$ rad m$^{-2}$, which corresponds to
$\sigma_{\rm PA}\approx 1\fdg5$ and which is still small compared to the combined measurement error on $\Psi$,
i.e.~$\sigma_{\Psi}=5^\circ$. For a further discussion on the systematic effect of erroneous determinations of ${\rm PA}_0$ in statistical studies of pulsar spinÐvelocity alignment, the reader can refer to Section~\ref{sec:sys}.

Finally, the frequency-independent, systematic PA rotations, caused by phase offsets during the processing
chain, were removed by comparing the PAs of absolutely calibrated Parkes observations of 5 bright, polarised
pulsars (covering a wide range of parallactic angles) to those obtained with the Lovell, after the
aforementioned calibration steps. Similarly, for the Effelsberg observations, PSR B1929+10 was used as a
polarimetric calibrator. Further details concerning the new observations can be found in
Carr~(2007)\nocite{car07}.

The statistical investigation of the relative phases of the pulsar spin and velocity vectors is strongly
affected by the systematic uncertainties (see Section~\ref{sec:sys}). Therefore, a careful selection of the
best constrained PAs is necessary to minimise the ambiguity of our conclusions. One major systematic
uncertainty comes from the determination of the spin axis direction, which is most commonly derived from
pulsar polarisation profiles. Fig.~\ref{fig:fig1} shows all the integrated polarisation profiles, from the
aforementioned observations, that were used in this work.
The lower panel of each plot shows the total intensity profile (black) and the linear (red) and circular
(blue) polarisation profiles. The top panels show the profile of the polarisation position angle (PA) across
the pulse. In each of the plots a vertical, dashed black line is drawn at ${\rm PA}_0$, i.e. the position
angle of polarisation at the point of closest approach of the observer's LOS to the magnetic pole.

As well as the polarisation measurements, Fig.~\ref{fig:fig1} shows in green horizontal dashed lines, the
position angle of the proper motion, ${\rm PA}_v$, in order to provide a direct comparison between the
orientations of the spin and velocity vectors. The components of pulsar proper motions used to calculate ${\rm
PA}_v$ are the most accurate measurements available, taken from Hobbs et al (2005)\nocite{hllk05}, Johnston et
al.~(2005) and other sources referenced in Table~\ref{tab:tab1}. Given the measured values of ${\rm PA}_v$ and
${\rm PA}_0$, we can calculate the offset angle, $\Psi$, between the sky-projected spin-axis orientation and
velocity direction. Taking into account all the ambiguities mentioned above, $\Psi$ is given by $\Psi=({\rm
PA}_0-{\rm PA}_v) \mod 45^\circ$, with a standard deviation calculated from
$\sigma_\Psi=\sqrt{\sigma_0^2+\sigma_v^2}$, where $\sigma_0$ and $\sigma_v$ are the standard deviations of
${\rm PA}_0$ and ${\rm PA}_v$, respectively. Here and later in our analysis, we assume Gaussian statistics for
the PA distributions. All ${\rm PA}_v$, ${\rm PA}_0$ values and the corresponding $\Psi$ angles for the
pulsars in our sample are shown in Table~\ref{tab:tab1}.

In summary, we have tried to address the open question of whether pulsar spins and their respective velocities
are preferentially aligned, using 54, good-quality $({\rm PA}_0,{\rm PA}_v)$ value pairs, of which 16 came
from new measurements by Carr (2007)\nocite{car07}. To date, this is the largest sample of pulsar
spin--velocity offset angles considered in such investigation.

\section{Data Analysis} 
\label{sec:dataan}
If there is a SN-kick mechanism that promotes the alignment between the pulsar spin and the kick velocity,
then for a sample of pulsars that possess this property one should expect to see the majority of $\Psi$ values
distributed towards $0^\circ$. On the other hand, a uniform distribution of offset angles would indicate no
correlation. A complication comes from the fact that the gravitational potential of the Galaxy alters the
pulsar velocity vectors during the lifetime of a pulsar. Hence, such correlation would be stronger for young
pulsars (e.g.~$\lesssim 1,000$ kyr) and vanish for older pulsars (e.g.~$\gtrsim 10,000$ kyr). The
investigation of spin--velocity alignment as a function of age is an interesting problem in its own right, but
it critically depends on the reliability of pulsar ages and distances. In the present paper, we do not make a
distinction between different pulsar age groups or consider the effect of the Galactic gravitational potential
on pulsar velocities; a detailed study of the above problem will be presented in a separate, follow-up
publication.


A simple way of checking whether the conclusions derived from the above studies hold for our selected sample
of pulsars is by quantifying how different the distribution of $\Psi$ is from a uniform distribution, using a
statistical test for similarity, like the KS test. Our limited sample of $N=54$ values of $\Psi$ can be simply
binned in steps of 5$^\circ$, between 0 and 45$^\circ$, to produce a discrete distribution of the offset
angles. However, as is clear from Table~\ref{tab:tab1}, the errors on ${\rm PA}_{\rm 0}$ and ${\rm PA}_v$ are
significant and have to be taken into account in quantifying the uniformity of the distribution.
Statistically speaking, a parent population of pulsars, part of which is our sample, would produce a
distribution of $\Psi$ that is the convolution of the individual Gaussian distributions of each pulsar in our
sample, each having a standard deviation equal to the measured one. A numerical way of approximating the
parent distribution using the available data is by performing a large number of Monte Carlo (MC) iterations,
allowing in each random instance the values of ${\rm PA}_0$ and ${\rm PA}_v$ to be distributed normally in
$N({\rm PA}_0,\sigma_0)$ and $N({\rm PA}_v,\sigma_v)$, respectively. The resulting values of
$\Psi$ from each iteration are then binned in $\Psi\in[0^\circ,45^\circ]$. Following this recipe, we generated
$10^6$ histograms of $\Psi$, which we then averaged to produce the final, aggregate distribution of $\Psi$;
the distribution is shown in Fig.~\ref{fig:psihist}a. This MC procedure shuffles the data positions between
bins of $\Psi$, according to the $\sigma_\Psi$ of each pulsar, which in turn causes the fluctuation of the bin
heights, $n_k$, between iterations. Therefore, a good measure of the statistical error on the $k$th bin,
$\sigma_k$, in the aggregate distribution is the standard deviation of the bin's height across all MC
iterations: i.e.~$\sigma_k=\sqrt{\langle n^2_k\rangle_{\rm MC}-\langle n_k\rangle^2_{\rm MC}}$.

We examined the uniformity of the aggregate distribution of $\Psi$ that was generated from the sample of
Table~\ref{tab:tab1} using the described procedure. At first glance, the distribution appears non-uniform with
$\approx 70$\% of the $\Psi$ values distributed below the median ($=22\fdg5$). As mentioned earlier, a
suitable test for quantifying the similarity of the aggregate distribution to a uniform one is the KS test.
In each of the iterations of the MC procedure, we performed a KS test between the randomly generated
distribution of $\Psi$ and a uniform probability function, $p(\Psi)=1/45$ deg$^{-1}$. Thus, we generated
$10^6$ values of the KS statistic, $D$. As is shown in Fig.~\ref{fig:psihist}b, the values of the statistic $D$ are distributed
in a Gaussian fashion, which justifies choosing the mean, $\bar{D}$, and the standard deviation, $\sigma_D$,
as the most probable value and its 68\% CLs, respectively. Based on these values, the
probability of rejecting uniformity under the KS test was calculated from $\bar{D}$, i.e.~$p(\bar{D})$,
and the probabilities corresponding to the 68\% CLs, from $\bar{D}\pm\sigma_D$,
i.e.~$p(\bar{D}\mp \sigma_D)$\footnote{Note the inverted signs, as we are interested in the probability of
rejecting uniformity.}.
These values are shown in the legend of Fig.~\ref{fig:psihist}b.
Hence, the apparent non-uniformity of the $\Psi$ distribution is confirmed by the KS test, which rejects uniformity at
the $99.8^{-1.2}_{+0.2}\%$ level, i.e.~at roughly $3\sigma$.

\begin{figure*} 
\vspace*{10pt}
\includegraphics[width=1\textwidth]{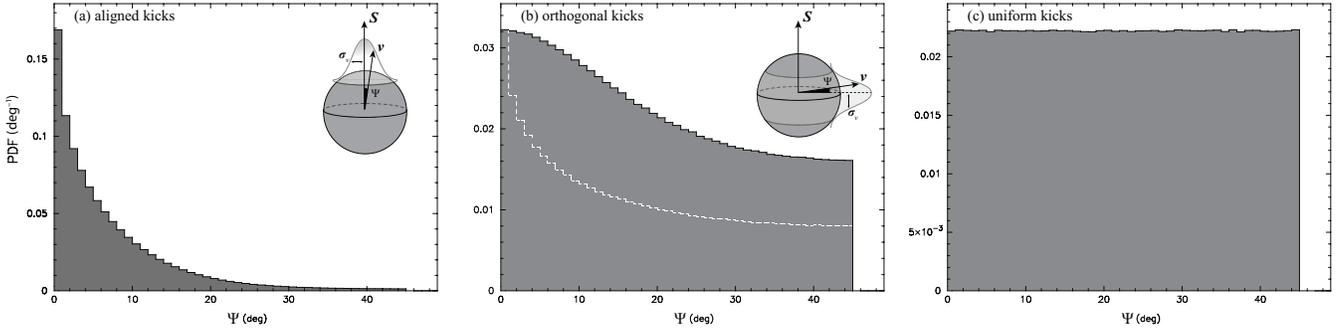}
\caption{\label{fig:simpsi} 
Simulated distributions of $\Psi=({\rm PA}_0-{\rm PA}_v) \mod 45^\circ$, for three cases of alignment
between the pulsar velocity, $\boldsymbol{v}$, and its spin axis, $\boldsymbol{S}$: (a) the distribution of
$\boldsymbol{v}$ is aligned with $\boldsymbol{S}$ to within $\sigma_v$; (b) the distribution of
$\boldsymbol{v}$ is orthogonal to $\boldsymbol{S}$ to within $\sigma_v$; and (c) the distribution of
$\boldsymbol{v}$ is uniform on a sphere. The geometric details of the simulation are explained under
Fig.~\ref{fig:losconfig}. For case {\em b}, the $\Psi$ distribution for $\sigma_v=0^\circ$ is shown with a
dashed white line. }
\end{figure*}

In addition to plotting the histogram of the wrapped angle $\Psi$, we wished to examine the raw, 2-dimensional
distribution of PAs, i.e.~${\rm PA}_v$ and ${\rm PA}_0$, without considering the possibility of orthogonal
emission but still allowing only the polarisation-measurement ambiguity. Since we are only interested in the
relative direction of the spin and velocity axes, we set all measured ${\rm PA}_0$ values to $0^\circ$ and
re-calculated ${\rm PA}_v$ as ${\rm PA}_v-{\rm PA}_0$, which was forced to be between $-90^\circ$ and
$90^\circ$. By randomising, as before, both angles within their respective statistical errors (and wrapping
their difference modulo $90^\circ$, as necessary), we generated a large number of $({\rm PA}_0,{\rm
PA}_v)$ pairs. The 2-dimensional probability density function (PDF) of the generated distribution is shown in
Fig.~\ref{fig:psihist}c. The PDF clearly shows an appreciable concentration of values around
$(0^\circ,0^\circ)$ and near $(0^\circ,\pm 90^\circ)$: this picture is consistent with the 1-dimensional
distribution of $\Psi$, as all the values inside the most significant contours ($p>0.0015$) would fall into
the bins below the mid-range of $\Psi$, if folded back into the $0^\circ-45^\circ$ interval. Also, at first glance, the central
island near $0^\circ$ appears the most significant of the three, having a central probability contour
($p\gtrsim 0.002$) that is more extended than both the one near $-90^\circ$ and that near $90^\circ$. However,
it is difficult to visually compare the likelihood of the aligned ($\sim 0^\circ$) with that of the orthogonal
($\sim \pm 90^\circ$) configurations, only based on the data, without modelling their expected distributions.
A quantitative comparison between these two configurations, and for various ${\rm PA}_v$ distributions, is
presented in the following Section.

In conclusion, if we take the PDF of Fig.~\ref{fig:psihist}c at face value, it appears that the concentration
of $\Psi$ values below $22\fdg5$ in the histogram of Fig.~\ref{fig:psihist}a is the result of not only nearly
parallel but also nearly orthogonal spin--velocity configurations. In particular, if we consider that having
${\rm PA}_v-{\rm PA}_0 = 90^\circ$ and $-90^\circ$ is essentially the same configuration, then both aligned
and orthogonal configurations appear to have comparable probability of occurrence, based on the PDF shown.
Equally interesting is the fact that PA offsets in the ranges $[-60^\circ$,$-30^\circ]$ and
$[30^\circ$,$60^\circ]$ appear very unlikely according to the PDF, which confirms that the observed alignment
is robust. However, as was conclusively shown with radio and X-ray observations of the Vela pulsar (Ng \&
Romani 2004\nocite{nr04}; Ng \& Romani 2007\nocite{nr07}), configurations that appear orthogonal, due to
orthogonal-mode emission, may actually be aligned. In such case, it is possible that Fig.~\ref{fig:psihist}c
shows a distribution of only aligned configurations, some of which appear orthogonal; while of course the
opposite is also possible.

\begin{figure*} 
\vspace*{10pt}
\includegraphics[width=0.57\textwidth]{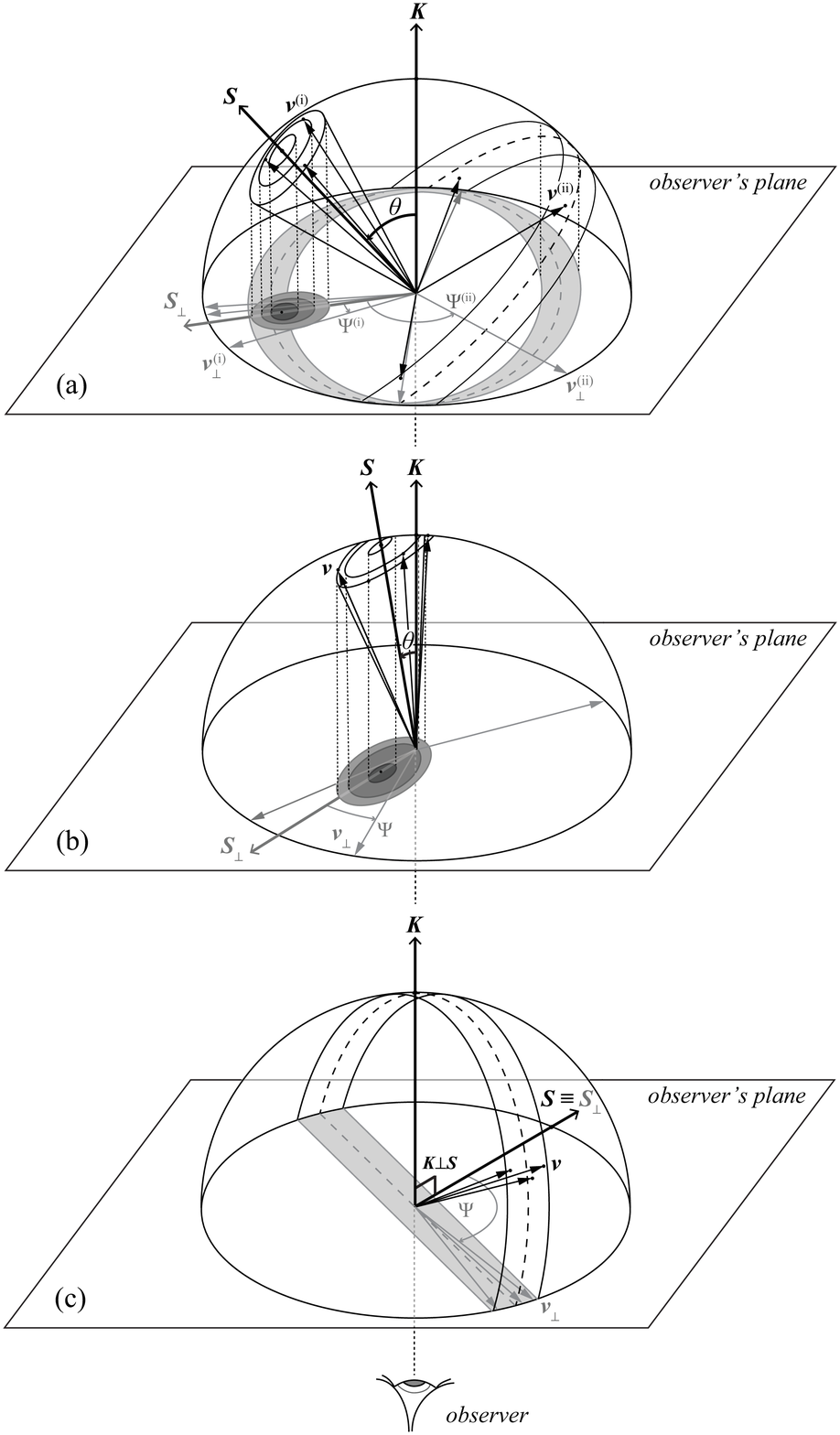}
\caption{\label{fig:losconfig} 
Schematics of the geometry used in our spin--velocity simulation code: (a) the top schematic shows the general
case, where the observer's LOS, $\boldsymbol{K}$, and the spin-axis vector, $\boldsymbol{S}$, form an
arbitrary angle, $\theta=\boldsymbol{K}\angle\boldsymbol{S}$; (b) the middle schematic shows the special case,
where $\boldsymbol{K}\angle\boldsymbol{S}$ is small, i.e.~$\theta\ll 45^\circ$; (c)
finally, the bottom schematic shows another special case, where $\theta=90^\circ$.
Assuming a uniform distribution of the observer's LOS and the spin-axis directions over $4\uppi$, case (b) is
less probable than case (c). In all schematics, the real, three-dimensional vectors for the pulsar spin and a
few exemplifying velocity vectors, $\boldsymbol{v}$, are represented with solid, black vectors; the solid,
grey vectors correspond to the projections of the three-dimensional vectors onto the plane perpendicular to
$\boldsymbol{K}$ (observer's plane). Plot (a) showcases two configurations of the pulsar's velocity vectors
with respect to $\boldsymbol{K}$ and $\boldsymbol{S}$, designated with the superscripts (i) and (ii): those of
(i) are part of a distribution of velocities about the spin axis, within a solid angle surrounding
$\boldsymbol{S}$ (probability density contours of increasing angular separation from $\boldsymbol{S}$ are also
shown); those of (ii) are part of a distribution of velocities along the equator (dashed black line), within
an equatorial band (delimited by solid black lines). The projection of the three-dimensional distributions of
velocities onto the observer's plane, for both configurations, is shown as two-dimensional, grey-shaded
probability density contours (for configuration (ii), the projection of the distribution that is below the
observer's plane is also drawn). On the observer's plane, the angles $\Psi$ between the projected spin and
velocity vectors, $\boldsymbol{S}_\perp$ and $\boldsymbol{v}_\perp$, respectively, are also shown. Using the
same conventions as in (a), plot (b) shows the case where the velocities are distributed about the pulsar's
spin axis and are also roughly aligned with $\boldsymbol{K}$. Finally, plot (c) shows the special case where
the velocities are distributed along the equatorial plane of the pulsar and the pulsar's spin axis lies in the
observer's plane; in this case, the projection of the velocity distribution degenerates into a diametral band,
yielding high probability for $\Psi\sim 90^\circ$.}
\end{figure*}

\section{Discussion}
\label{sec:discuss}

\subsection{Toy-Model Simulations}
\label{sec:toymodel}

\subsubsection{Model description \& geometry}
In order to further quantify the margins of the process that could potentially lead to the observed
non-uniformity in the distribution of $\Psi$, we performed numerical, toy-model simulations. We simulated a
high number of pulsar-velocity directions over $4\uppi$ rad$^2$, assuming a parent process that results in (a) a
peri-axial distribution of velocities (i.e.~case of aligned kicks), (b) a distribution of velocities nearly
aligned with the equatorial plane (i.e.~orthogonal kicks) and (c) a uniform distribution of velocities with
no preferred direction. Furthermore, in order to simulate the random orientations of pulsar spins with respect
to the observer, we allowed the observer's LOS to be distributed uniformly over the entire sky but kept the
pulsar spin-axis orientation fixed. After projecting the spin and velocity directions onto the observer's
plane (perpendicular to the LOS), we folded the resulting values of $\Psi$ in the $0^\circ-45^\circ$
interval (as was done with the real data). The character of the resulting distributions of $\Psi$ for those three cases can
be seen in Fig.~\ref{fig:simpsi}, where we arbitrarily used a Gaussian standard deviation of $10^\circ$ for
the ${\rm PA}_v$ distributions (hereafter simply referred to as $\sigma_v$).


The observed distribution of $\Psi$ can be more clearly understood with the help of the schematics of
Fig.~\ref{fig:losconfig}, which show the three-dimensional geometry invoked in our simulation. It is clear
that a preferred alignment would lead to the $\Psi$ distribution of Fig.~\ref{fig:simpsi}a, where $>90\%$ of
the values are below $20^\circ$, and with the distribution practically vanishing towards $\Psi=45^\circ$. The
underlying geometry of this case is shown in Fig.~\ref{fig:losconfig}a, configuration (i). It is also worth
noting that uniformly distributing the observer's LOS over $4\uppi$ does not significantly affect the shape of
this distribution, because the only possibility of having $\Psi\gg 0^\circ$, in this case, would be for LOS that
are nearly parallel to $\boldsymbol{S}$, which is a configuration with low probability. Such a low-probability
case is shown in Fig.~\ref{fig:losconfig}b.

On the other hand, the orthogonal case results again in an uneven distribution with an excess of values below
$20^\circ$. As can be seen in Fig.~\ref{fig:losconfig}a, configuration (ii), the projection of the
three-dimensional velocity distribution onto the observer's plane, for an arbitrary angle between
$\boldsymbol{K}$ and $\boldsymbol{S}$, is an ellipse with its minor axis aligned with $\boldsymbol{S}_\perp$
(i.e.~the projection of $\boldsymbol{S}$ onto the observer's plane). If we allow the distribution of
$\boldsymbol{v}$ to have a width around the equator, then the projection of this equatorial band onto the
observer's plane is an elliptical ring-shaped area --- formed by the superposition of ellipses within the
equatorial band --- that is broadest along the direction of $\boldsymbol{S}_\perp$ and whose width diminishes
perpendicular to it. At the extreme ends of the possible $\theta=\boldsymbol{K}\angle\boldsymbol{S}$
configurations, that projection degenerates into either a diametral band (for $\theta=90^\circ$; see
Fig.~\ref{fig:losconfig}c) or a circular band (for $\theta=0^\circ$ or 180$^\circ$; not shown). In the former
case, the most probable values of $\Psi$ are $\sim 90^\circ$, whereas in the latter case $\Psi$ can take all
values in $[0^\circ,180^\circ]$ with equal probability. Crucially, as was mentioned in the previous paragraph,
the probability of having $\theta=0^\circ$ (or 180$^\circ$) is significantly lower than that for
$\theta=90^\circ$, in a uniform distribution of LOS and $\boldsymbol{S}$ directions. Therefore, after folding
all $\Psi$ angles modulo $45^\circ$, there will be an excess of angles near $\Psi\sim 0^\circ$ compared to
those near $45^\circ$. And this is true independently of any dispersion of the velocity directions around the
equator, as is clearly shown with the dashed white line in Fig.~\ref{fig:simpsi}b, where no deviation of the
velocity direction from orthogonality was assumed.

Finally, as expected, the case of uniformly distributed velocities results in a uniform distribution of
$\Psi$.

\begin{figure*} 
\vspace*{10pt}
\includegraphics[width=0.75\textwidth]{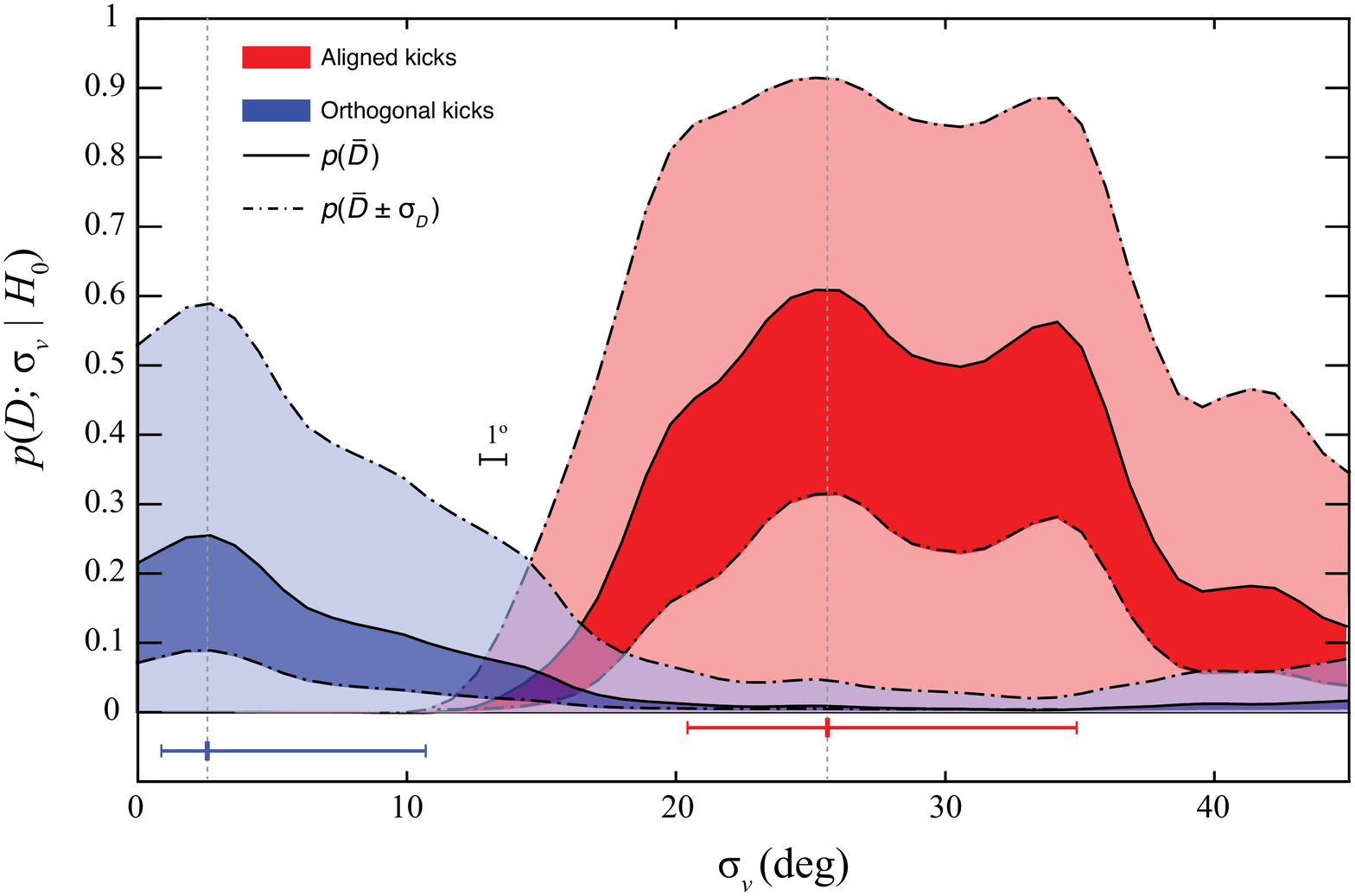}
\caption{\label{fig:simpsiPDFs} The PDFs of $\sigma_v$, in the case of aligned (red-shaded areas) and
orthogonal (blue-shaded areas) kicks, from the comparison (under a two-sample KS test) of the data
distribution of $\Psi$ for all pulsars in our sample with the simulated histograms of Fig.~\ref{fig:simpsi},
for a range of trial $\sigma_v$ values. The central PDFs, for each case, are delimited with solid black lines
and correspond to the mean value of the KS statistic, $\bar{D}$, from our MC simulation, for each $\sigma_v$:
i.e.~$p(\bar{D})$. The PDFs delineated with dash-dotted black lines correspond to the 68\% confidence limits
of the KS statistic, $\bar{D}\pm\sigma_D$: i.e.~$p(\bar{D}\pm\sigma_{D})$. The correspondingly coloured error bars along the
horizontal axis denote the median and $1\sigma$ asymmetric errors on $\sigma_v$, based on the central PDFs of
the alignment and orthogonal case. Finally, the black error bar corresponds to the step in $\sigma_v$ used in
the simulated $\Psi$ distributions.}
\end{figure*}

\subsubsection{Fits to the data}
\label{ssubsec:datfits}
The significant departure from a flat distribution of the observed distribution of $\Psi$ gives us the
opportunity to investigate the properties of the parent distribution of 3D velocities by comparing it with our
toy-model simulation. In order to produce a quantifiable measure of similarity, we performed a two-sample KS
test between the observed and simulated distributions of $\Psi$, for a range of $\sigma_v$ values. In addition
to the free parameter $\sigma_v$, we added a fixed amount of observational error on $\Psi$, roughly equal to
the average value measured in our data set: i.e.~$\sigma_{\Psi}=10^\circ$. Thereafter, we generated 100
simulated distributions of $\Psi$ for $\sigma_v\in[0^\circ,45^\circ]$, each containing 100,000 values. For
each of the simulated distributions, we generated 10,000 MC data sets by normally distributing the measured
$\Psi$ values according to their standard deviations, similarly to the procedure done for
Fig.~\ref{fig:psihist}. 

For each $\sigma_v$, we performed a two-sample KS test between the simulated distribution $\Psi(\sigma_v)$ and
each of the 10,000 data sets. Each of these tests produces a KS statistic, $D$; the whole procedure produces a
distribution of 10,000 values of $D$, similar to that of Fig.~\ref{fig:psihist}b --- which was derived from
the one-sample KS test between $p(\Psi)=1/45$ deg$^{-1}$ and a large number of data sets. From that
distribution, we calculated the mean KS statistic, $\bar{D}$, and its 68\% CLs, $\bar{D} \pm \sigma_D$. Finally, the
corresponding probability of rejecting the null hypothesis under the KS test was calculated from $\bar{D}$ and
$\bar{D} \pm \sigma_D$:
i.e.~$p(\bar{D}; \sigma_v | H_0)$ and $p(\bar{D}\pm \sigma_D; \sigma_v | H_0)$, respectively. The resulting
PDFs for the cases of aligned and orthogonal kicks are shown in Fig.~\ref{fig:simpsiPDFs}. As well as the
central PDFs, which correspond to the $p(\bar{D}; \sigma_v | H_0)$ values (black lines), we have also plotted
those corresponding to the 68\% CLs on $D$ (dashed-dotted lines), as was explained above.
For clarity, the areas between $p=0$ and each of the PDF functions shown (with solid and dashed-dotted lines)
have been shaded with dark-red and dark-blue colours, for the central PDFs of the aligned and orthogonal case,
respectively, and with light-red and light-blue colours, for the PDFs of the aligned and orthogonal case,
respectively, corresponding to the 68\% limits on $D$. An immediately evident feature of the generated PDFs is
that they show clear global maxima for certain values of $\sigma_v$. For the case of alignment, the central
PDF gives a 68\% confidence interval around the maximum of $\sigma_v=26^{+9}_{-6}$ deg.
Interestingly, the PDF in the orthogonal case also exhibits a clear peak, but in that case the peak
probability is less than half that of the case of alignment ($p_{\perp}\approx 0.25^{+0.35}_{-0.15}$); the
latter case results in a probability of $60\pm30\%$ that the observed and simulated distributions are the
same. For the central PDF, the case of orthogonal kicks gives a peak at $\sigma_v=3^{+8}_{-2}$ deg.

\subsubsection{Conclusions}

In conclusion, the comparison between our toy-model simulations and the data have revealed a preference
towards alignment between velocity and spin, if $\sigma_v\approx30^\circ$ --- although it should be stressed
that in both cases the PDFs are very broad and irregular, and therefore $\sigma_v$ is not very well
constrained. This result also reveals that, in fact, the central island of Fig.~\ref{fig:psihist}c is more
significant than those close to $\pm 90^\circ$. The most probable ranges of $\sigma_v$ obtained from our
analysis can be compared with those derived from the simulations of KPP using data from Ng \& Romani (2007)
and Rankin (2007; see Introduction). For the case of alignment, our range of $\sigma_v$ values implies a much
larger spread of kick angles than the narrow distribution of kick angles using the data by Rankin
($\lesssim5^\circ-10^\circ$). On the other hand, the case of orthogonal kicks, which is favoured by scenarios
of binary disruption, is much more consistent with the above, narrow range. Interestingly, according to KPP,
the data by Rankin can {\em hardly} be fitted without including binary progenitors in the kick model.
Furthermore, as was mentioned in the introduction, the data of Ng \& Romani can be fitted by a much broader
distribution of kick angles ($\lesssim20^\circ$). This result is more consistent with our case of alignment,
although we find a somewhat broader spread of angles ($\lesssim20^\circ-35^\circ$).

Ultimately, our analysis shows that both aligned and orthogonal kicks appear probable and that it is difficult
to confidently rule out one of the modes in favour of the other. Hence, we are forced to consider the
possibility that an ensemble of kick mechanisms may be in operation in the SN population, resulting in a
mixture of aligned and orthogonal kicks. Also, the significant fraction of pulsars that are born in binary
systems complicates things further: our analysis cannot discriminate between spin--velocity angles that are
the result of a binary motion superimposed on the SN birth kick. Furthermore, as will be discussed in a follow
up paper, the role of the Galactic gravitational potential is also an important one and can
drastically distort any spin--velocity correlation that may have been originally in place.

As a final remark, it should be stated that our simulation has not considered the pulsar-beam orientation with
respect to the observer, i.e.~the detectability of the simulated pulsars. In order to investigate whether beam
geometry has an effect on the simulated distributions, we reran our simulation, this time excluding pulsars
whose beam did not intersect with the observer's LOS. This condition demands certain assumptions about the
angular radius of the pulsar emission cone, $\rho$, and the distribution of the magnetic inclination,
$\alpha$. Based on the work of Lyne \& Manchester (1988)\nocite{lm88}, we used $\rho=6\fdg5 P^{-1/3}$,
which corresponds to a median of $\bar{\rho}\approx 5^\circ$, for our pulsar sample; hence, we chose
$\rho=5^\circ$ as the beam width. Furthermore, there are a number of PDFs for $\alpha$, $p(\alpha)$, in the
literature (see e.g.~Gil \& Han 1996\nocite{gh96}): the most prominent ones are a uniform distribution in
$\alpha$ and in $\cos\alpha$, i.e.~$p(\alpha)=2/\pi$ and $p(\alpha)=\sin\alpha$, respectively --- where
$\alpha\in [0,\uppi/2]$; the latter of the two distributions is just a uniform distribution of the beam
directions over $4\uppi$ ({\em cf.} the same distribution was assumed for $\boldsymbol{K}$).
Other, more complicated distributions exist, but, in general, the limited sample of available measurements
combined with the aforementioned issues with RVM fitting (and other methods of obtaining $\alpha$; e.g.~Rankin
1990\nocite{ran90}), as well as our lack of understanding of the evolution of $\alpha$ with pulsar age
(e.g.~Tauris \& Manchester 1998\nocite{tm98}; Weltevrede \& Johnston 2008\nocite{wj08}), does not warrant in our opinion such complicated descriptions.
As with the observer's LOS, we randomly chose $\alpha$ from the uniform distribution,
accepting only configurations for which $\alpha-\rho \leq \theta\leq \alpha+\rho$ or $(\pi-\alpha)-\rho \leq
\theta\leq (\pi-\alpha)+\rho$. As before, $\sigma_v$ was kept fixed to $10^\circ$.
As is expected for a uniform distribution in $\alpha$, for each randomly generated configuration of
$\boldsymbol{K}$ and $\boldsymbol{v}$, only a certain fraction of beam orientations are accepted; but that
fraction does not depend on $\alpha$ and, hence, the final $\Psi$ distributions remain unchanged, in this
case. On the other hand, choosing a uniform distribution in $\cos\alpha$ resulted in practically the same
$\Psi$ distribution for the aligned case. For the orthogonal case, a uniform $\cos\alpha$ distribution
resulted in a slightly steeper distribution for $\Psi$ compared to the one shown in Fig.~\ref{fig:simpsi}b.
However, by increasing $\sigma_v$ by 20\%, we were able to match the two distributions, meaning that within
20\% adding pulsar detectability does not affect the simulated distributions. In summary, the magnitude
of the errors on the determined values of $\sigma_v$, for the orthogonal case, ranges from 60\% to 260\%; this
dominates over the bias from excluding detectability. And since there is currently no consensus in the
literature as to which distribution of $\alpha$ gives the best description, we decided not to take this into
account in our simulations.


\subsection{Systematics Inspection}
\label{sec:sys}

The above treatment neglects possible systematic errors, which can have a significant effect on the
distribution of Fig.~\ref{fig:psihist}. It is therefore important to test the robustness of the observed
correlation between the spin and velocity position angles against systematic departures from the data set of
Table~\ref{tab:tab1}. The following paragraphs consider two, potentially significant sources of bias: the
choice of ${\rm PA}_0$ and the influence of the J05 subset on the degree of correlation.  

\begin{figure*} 
\vspace*{10pt}
\includegraphics[width=1\textwidth]{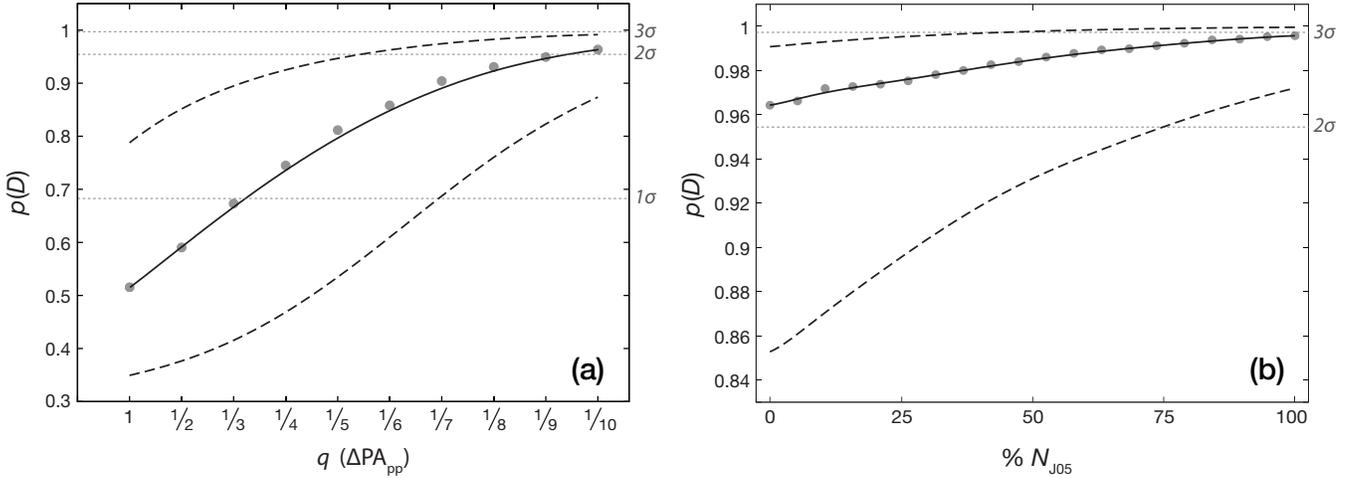}
\caption{\label{fig:dtests} Probability of rejection of uniformity, under the KS test, as a function of (a)
departure of ${\rm PA}_0$ from the published values in units of the PA profile's peak-to-peak difference (as
shown in Fig.~\ref{fig:fig1}), and (b) fraction of values from Johnston et al.~(2005) in the data. In both plots,
the dashed black lines correspond to the $1\sigma$ confidence interval around the probability of the mean KS
statistic. The dotted grey lines mark the 68, 95.4 and 99.6\% CLs of a normal distribution.}
\end{figure*}

\subsubsection{The choice of ${\rm PA}_0$}
\begin{figure*} 
\vspace*{10pt}
\includegraphics[width=1\textwidth]{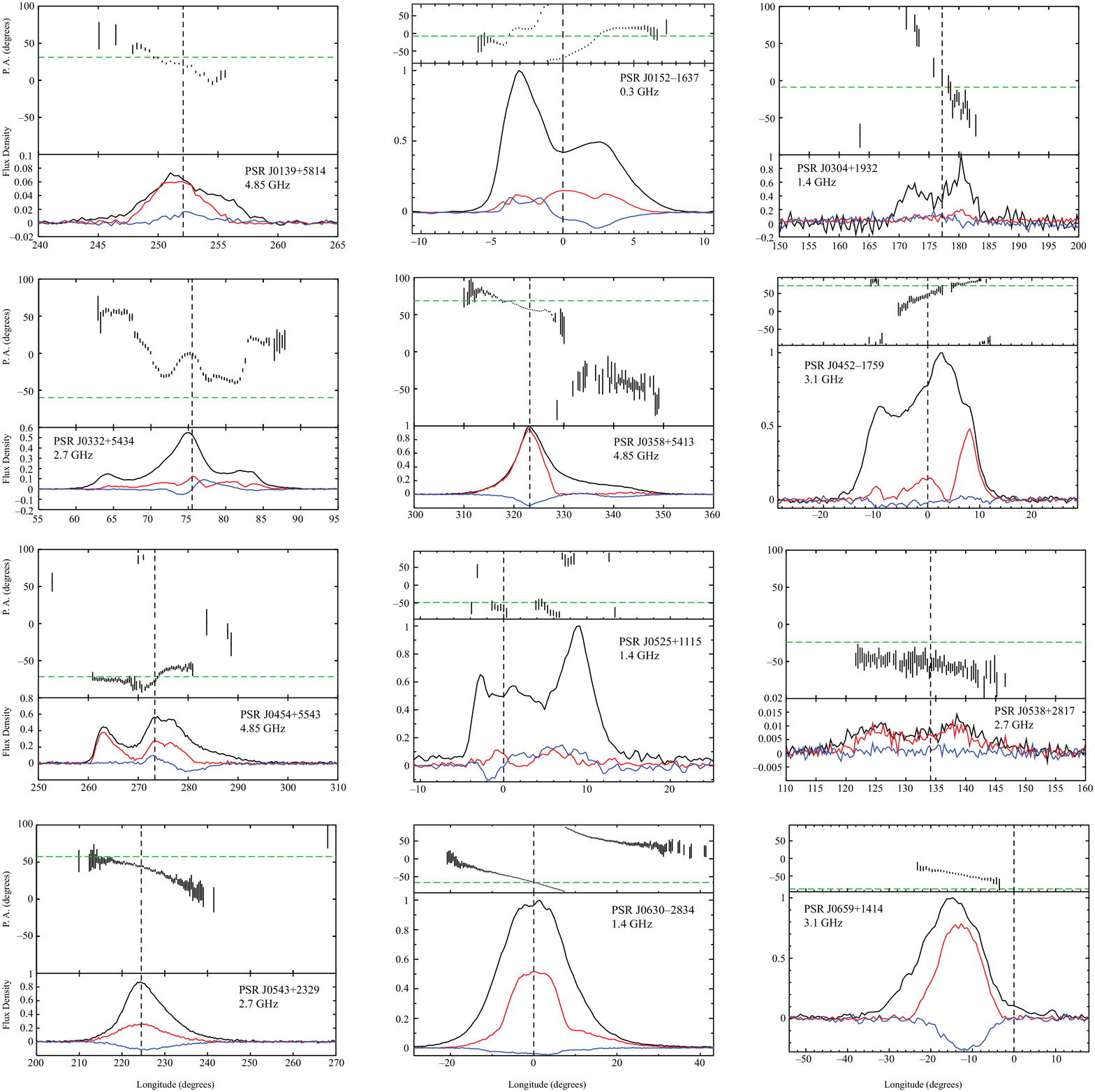}
\caption{\label{fig:fig1} Polarisation and position-angle (PA) profiles for 51 pulsars taken from Johnston et
al.~(2005, 2007) and Carr (2007). The lower panel of each plot shows the total intensity profile (black) and
the linear (red) and circular (blue) polarisation profiles. The top panels show the profile of the PA across
the pulse, for all phase bins with S/N of the linear polarisation greater than 3,
i.e.~$L/\sigma_I>3$. In each of the plots a vertical, dashed black line is drawn at ${\rm PA}_0$, i.e.~the
position angle of polarisation at the point of closest approach of the observer's LOS to the magnetic pole.
Also shown in these plots is the position angle of the proper motion vector (dashed, green line).}
\end{figure*}

As was stated in Section~\ref{subsec:prevwork}, a potential bias in our study of pulsar-spin alignment using
pulsar polarisation is the choice of ${\rm PA}_0$. The determination of the reference PAs that were used in
this paper was based on a number of criteria employed by the respective investigators (see
Section~\ref{sec:observations}). However, due to the limited statistics, the results of such analysis are
potentially sensitive to erroneous determinations of ${\rm PA}_0$. For example, the PA profile of PSR
J0630$-$2834 sweeps across the entire range of possible angles, i.e.~$\Delta{\rm PA}_{\rm
pp}\approx180^\circ$; here, $\Delta{\rm PA}_{\rm pp}$ is defined as the peak-to-peak difference of the PA
profile for across all phases with $L/\sigma_I>3$, i.e.~S/N in $L$ greater that 3 (as shown in
Fig.~\ref{fig:fig1}). Clearly, the decision of ${\rm PA}_0$ for this pulsar will have a significant impact on
whether it shows spin--velocity alignment or not; on the other hand, the characteristic RVM shape of the PA
profile assists an easier determination of the inflexion point. In contrast, PSR J2013+3845 possesses a flat
PA profile ($\Delta{\rm PA}_{\rm pp}\approx30^\circ$), so an error on the choice of ${\rm PA}_0$ would not
significantly alter the result --- but whether the relative constant PA across the profile corresponds to the
PA at the minimum approach to the magnetic pole or not is unknown. In order to test how sensitive our data
sample is to the choice of ${\rm PA}_0$, we generated a high number of data sets, choosing each time a
different ${\rm PA}_0$ for each pulsar, from $N(PA_0,q\Delta{\rm PA}_{pp})$, where $q$ ranged from 1
(corresponding to a standard deviation equal to the entire range of possible PAs) to $\nicefrac{1}{10}$
(corresponding to a standard deviation equal to 10\% of $\Delta{\rm PA}_{\rm pp}$). The value of $\Delta{\rm
PA}_{\rm pp}$ varied from pulsar to pulsar according to the PA profiles of Fig.~\ref{fig:fig1}. Again, for
each of the above data sets we performed the usual MC procedure, in order to estimate the mean values of the
KS statistic. The standard deviations of the randomly generated values of ${\rm PA}_0$, from the previous
step, were kept equal to those of the published PAs, i.e.~$\sigma=\sigma_0$, independently of the actual value
that a PA may have had at those same phases (PAs near the edges of the profiles have in general higher
$\sigma$ due to the lower S/N). The probabilities corresponding to the median value of the KS
distribution of means and its 68\% CLs are shown as a function of $q$ in Fig.~\ref{fig:dtests}a.

The most apparent feature of Fig.~\ref{fig:dtests}a is the evident trend towards higher probabilities for
decreasing errors on ${\rm PA}_0$. It is true, however, that the degree of spin-velocity
alignment appears very sensitive to the magnitude of the departure from the published values of ${\rm PA}_0$:
standard deviations of $>30\%$ of the $\Delta{\rm PA}_{\rm pp}$ result in practically no alignment ($p<0.8$),
whereas rejection of $H_0$ at the $2\sigma$ level is only achieved for $q\lesssim0.1$. The immediate
conclusion from this test is that the observed alignment remains robust to within a 10\% systematic error on
the choice of PA$_0$ (for a $2\sigma$ rejection of $H_0$). As was suggested above, beyond the impact of
systematic errors induced by human choice, such tests are unable to account for the intrinsic differences in
the polarisation geometry between pulsars: thus, we have accepted at face value that the best determined ${\rm
PA}_0$ corresponds to the polarisation position angle at the minimum approach to the magnetic pole, which
might not be the case, even for flat PA profiles, like that of PSR J2013+3845.

\subsubsection{The influence of subsets}
Arguably, the best, statistically significant observational support for pulsar spin--velocity alignment has
come so far from the data of J05 --- including the re-examination of those data by Rankin (2007). Follow up
work, e.g.~by Johnston et al.~(2007), has provided a less convincing picture --- although mitigating arguments
were presented to explain this. Our statistical analysis included $N_{\rm J05}=19$ pulsars from the data set
of J05, and the degree of observed alignment must therefore have been influenced by the properties of that
subset. In order to gauge how independent our analysis is compared to that of J05, we calculated the statistic
$D$ and it associated probability $p(D)$, as before, for data sets including a different number of pulsars
from J05, ranging from 0 to all 19; in each of the data sets, we kept the same total number of pulsars,
i.e.~$N-N_{\rm J05}=35$ pulsars, as a means to accommodate the case where no J05 pulsars were included and to
keep the samples statistically comparable. More specifically, the inclusion of different numbers of J05
pulsars in those data sets were performed in a bootstrapped manner, upon which a random sample (with
replacement) is chosen to be part of the $N$ pulsars in each data set. This process is re-iterated a large
number of times by means of calculating the distribution of $D$ for a particular $N_{\rm J05}$. As before,
each bootstrapped data set was subjected to the same MC procedure as for the original data set, in order to
calculate the mean value of the KS statistic. The probability of rejecting uniformity, $p(D)$, as a function
of $N_{\rm J05}$ is plotted in Fig.~\ref{fig:dtests}b. The error bars in that plot indicate the confidence
in the calculated central value of $p(D)$. Due to the positive skewness of the distribution of means from the
MC, we chose to plot the probability of the median and its $1\sigma$ asymmetric errors instead of that of the
mean.

As with the test on ${\rm PA}_0$, Fig.~\ref{fig:dtests}b shows a positive trend for increasing J05
fractions: i.e.~from $p(D)\approx 0.96$, when no J05 value is included in the data, to $p(D)>0.99$, when all
J05 values are included. In addition, it is clear that the confidence interval of $p(D)$ is significantly
decreased towards higher values of $N_{\rm J05}$. Both these facts advocate the significant impact of the J05
data set. It is true, however, that all bootstrapped data sets reject $H_0$ at a higher that 2$\sigma$ level,
even without including any J05 values --- although, it should be noted that the confidence interval in the
latter case extends to far below that level (i.e.~$p=0.96^{+0.03}_{-0.13}$).

\section{Summary}
We have revisited the long-standing problem of pulsar spin--velocity alignment with at least twice as many
pulsars as in other respective studies so far. A sample of 54 pulsars was selected from Parkes, Effelsberg and
Lovell observations, with accurate PA profiles and reliable proper motions, which was used to examine the
distribution of angles between the proper motion and spin-axis orientations. In addition, we included 3
pulsars for which the spin-axis orientation had been previously obtained by fitting the PWN torus surrounding
the pulsar. Allowing for the possibility of orthogonal emission and the intrinsic $\uppi$-ambiguity in
measurements of the polarisation vector, we simulated the distribution of $\Psi={\rm PA}_v-{\rm PA}_0
\mod 45^\circ$ based on the measurement errors on the individual PAs. The distribution of $\Psi$ appears
significantly non-uniform, with 70\% of the values residing below $20^\circ$ and the corresponding probability
of rejecting uniformity under the KS test being 99.8\%, which is a roughly 3$\sigma$ rejection of the case
that ${\rm PA}_v$ and ${\rm PA}_0$ are uncorrelated. Ignoring the possibility of orthogonal emission on the
other hand reveals that having spin and velocity orthogonal to each other is as likely as having them
aligned. Notably, the probability distribution exhibits clear minima for offsets between $30^\circ$ and
$60^\circ$, thus supporting the robustness of the alignment (or, indeed, orthogonality) between spin and
velocity.

The statistically well-defined shape of the distribution of $\Psi$, shown as a histogram in
Fig.~\ref{fig:psihist}, encouraged a comparison between the observed distribution and a number of simulated
distributions generated from spin-aligned and spin-orthogonal velocity distributions, having a range of
standard deviations. The statistical comparison showed that the alignment scenario is more probable than
spin--velocity orthogonality, for the observed distribution of angles, although the confidence intervals in
both cases are quite broad: a conclusion that can already be drawn from the PDF of Fig.~\ref{fig:psihist}.
Furthermore, this comparison allowed us to determine the spread that the kick-velocity distribution must have
to be consistent with the observations: we derived a $\approx 20^\circ-35^\circ$ spread for the case of
alignment. A much narrower, $<10^\circ$ spread was found for the orthogonal case, which is consistent with
previously published results that allow for binary kicks.

Finally, we checked the robustness of our conclusions against two systematic effects: we investigated (a) how
tolerant our conclusion is to departures from the best determined values of ${\rm PA}_0$ and (b) how much the
inclusion of the Johnston et al.~(2005) data set (the best statistically favourable evidence, yet) biases our
conclusions. In both cases, we found that observed alignment is robust: excluding the J05 data set does little
towards weakening the non-uniformity of the distribution of $\Psi$, although including it significantly
improves the $H_0$-rejection confidence; also, the observed alignment is favoured at $>95\%$ confidence to
within a 10\% excursion from the published ${\rm PA}_0$ values. It should be stressed that our robustness
analysis could only probe systematics induced by human choice: systematics intrinsic to the pulsar emission,
e.g.~the polarisation orientation relative to the magnetic field lines, cannot be assessed without further
information, such as X-ray imaging of PWN tori.

The strong evidence for a correlation between spin and velocity vectors in our sample strengthens the
conclusions of previous publications, but it also opens new questions about the physics behind the
correlation. The signature of a kick process that favours alignment at pulsar birth should not be traceable
for old pulsars ($\gtrsim 10,000$ kyr) whose velocities have been stochastically altered by the Galactic
gravitational potential. On the other hand, a sample of young pulsars ($\lesssim 1,000$ kyr; like the ones
selected in J05) should contain such a signature. Our sample did not make a distinction between those age
groups and yet a strong correlation was evident in the data ($\tau_{\rm c}\sim 1-100,000$ kyr). The
investigation of how robust the correlation is as a function of pulsar age is the subject of a follow-up
study.

\section{Acknowledgments}
The authors would like to thank the referee for a particularly constructive set of comments and suggestions. During the writing of this paper, the authors received invaluable advice from Drs.~Norbert Wex, Evan Keane, Lucas Guillemot and K.~J.~Lee, which helped to vastly improve this article. The Lovell Telescope is owned and operated by the University of Manchester as part of the Jodrell Bank Centre for Astrophysics with support from the Science and Technology Facilities Council of the United Kingdom. This work is partly based on observations with the 100-m telescope of the MPIfR (Max-Planck-
Institut f\"ur Radioastronomie) at Effelsberg.

\bibliography{journals,modrefs,psrrefs,crossrefs}

\vfill

\setcounter{figure}{1}

\begin{figure*} 
\vspace*{10pt}
\includegraphics[width=1\textwidth]{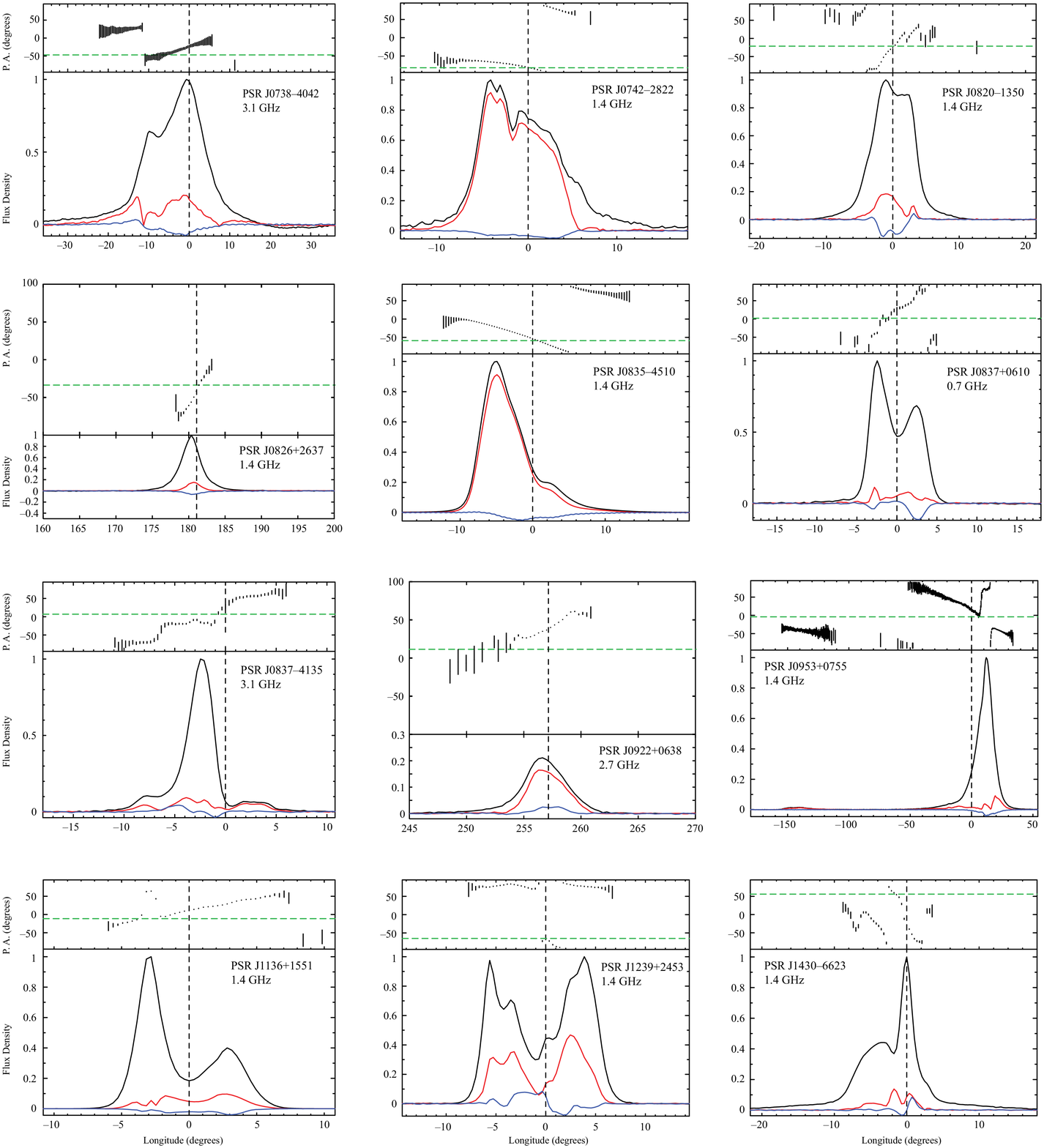}
\caption{\label{fig:fig2} Continued.}
\end{figure*}

\setcounter{figure}{1}

\begin{figure*} 
\vspace*{10pt}
\includegraphics[width=1\textwidth]{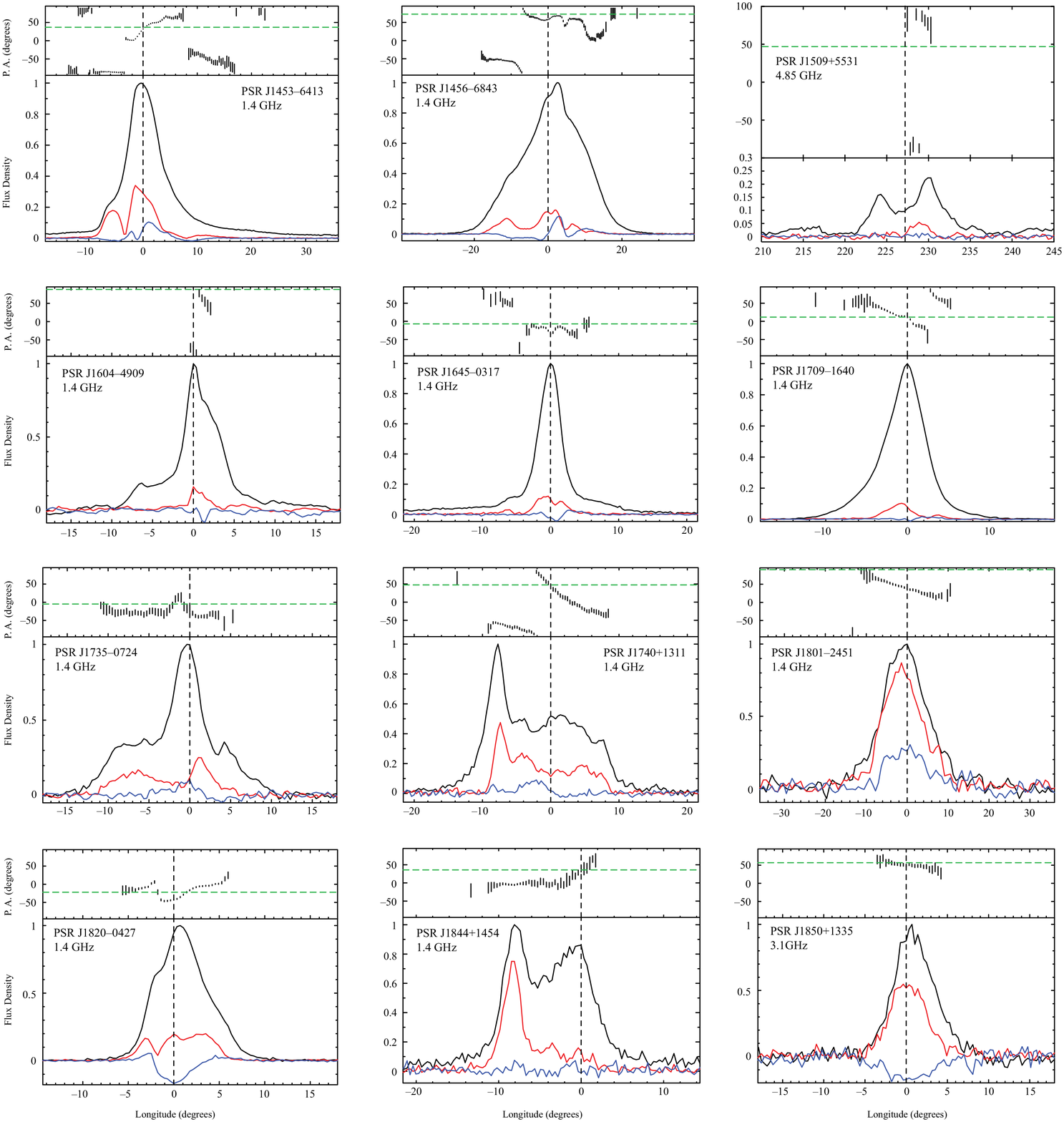}
\caption{\label{fig:fig3} Continued.}
\end{figure*}

\setcounter{figure}{1}

\begin{figure*} 
\vspace*{10pt}
\includegraphics[width=1\textwidth]{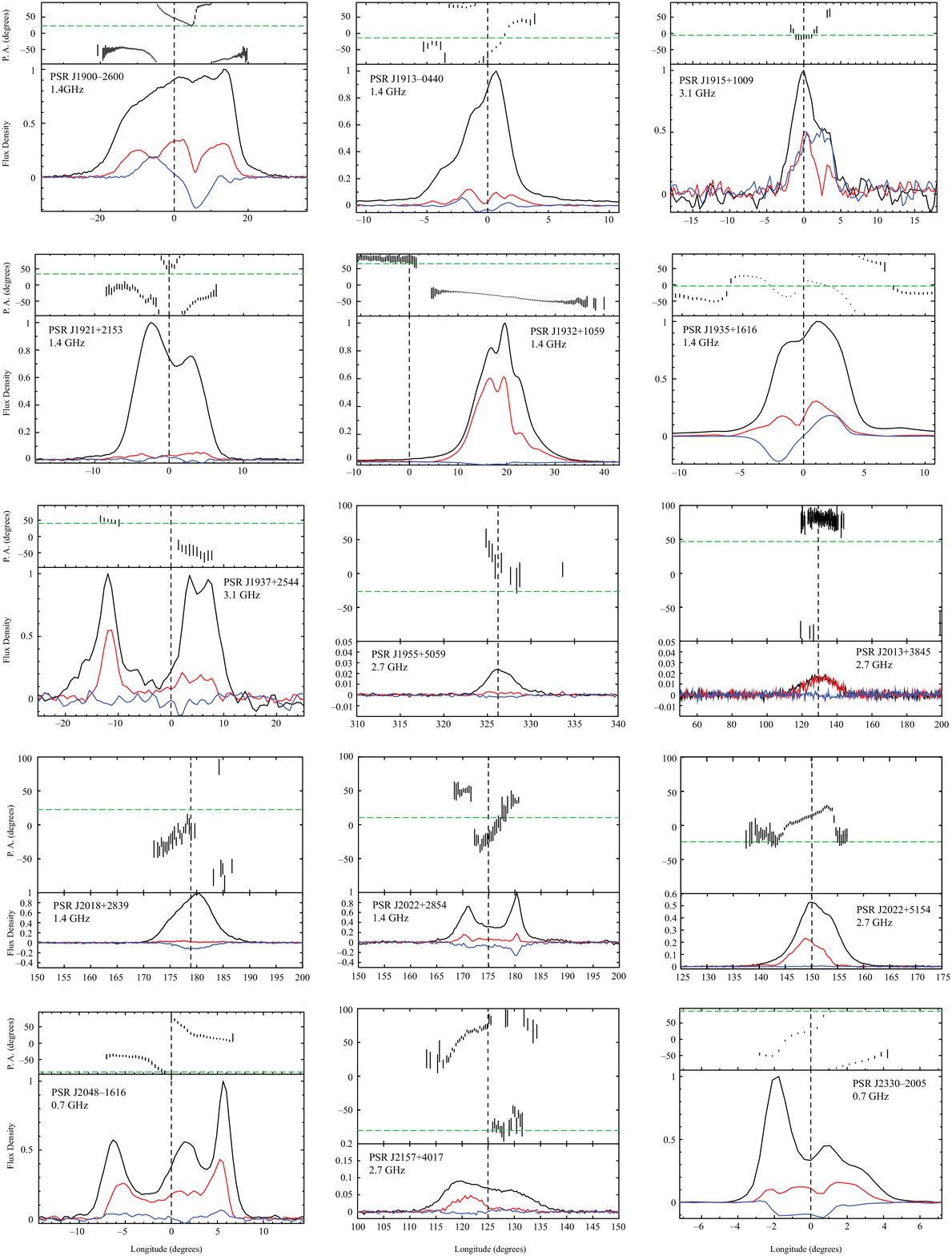}
\caption{\label{fig:fig4} Continued.}
\end{figure*}

\end{document}